# The influence of cations on the dipole moments of neighboring polar molecules

Imre Bakó, Dániel Csókás and István Mayer

Institute of Organic Chemistry, Research Centre for Natural Sciences,
H-1519 Budapest, P.O.Box 286, Hungary

Szilvia Pothoczki[1], László Pusztai[1,2]

[1]Wigner Research Centre for Physics, H-1121 Budapest, Konkoly Thege M. út 29-33., Hungary

[2]International Research Organization for Advanced Science and Technology (IROAST), Kumamoto University, 2-39-1 Kurokami, Chuo-ku, Kumamoto, 860-8555, Japan

## In Memory of István Mayer

Abstract

It is shown that the dipole moment of polar (water, methanol, formamide, acetone and acetonitrile) molecules in the neighborhood of a cation is increased primarily by polarization from the bare electrostatic charge of the cation, although the effective value of the latter is somewhat reduced by "back donation" of electrons from neighbouring polar molecules. In other words, the classical picture may be viewed as if a point charge slightly smaller than the nominal charge of the cation would be placed at the cation site. It was found that the geometrical arrangement of the polar molecules in the first solvation shell is such that their mutual polarization reduces the dipole moments of individual molecules, so that in some cases they become smaller than the dipole moment of the free protic or aprotic molecule. We conjecture that this behavior is essentially a manifestation of the Le Chatellier–Braun principle.

# Introduction

The uneven distribution of positive or negative charges in a molecule gives rise to a dipole moment. This dipole moment is uniquely defined, independently of the choice of coordinate system, if the total charge of the molecule is zero. The dipole moment (and also, other quantities that are sensitive to the deviation of the charge distribution from spherical symmetry) of an un-charged molecule can be calculated by defining the origin of the molecule in such way that the following equations holds true: $\sum_i^n \underline{r_i} = 0$, where $\underline{r_i}$ is the vector point from the centre of the coordinate system to the centre of the i-th centroid. The dipole moment of a molecule as a physical quantity can be defined using the equation below,

$$\underline{\mu} = \sum_i^n q_i \underline{r_i} + \int \rho(r)\underline{r}\, d^3r \qquad \text{Eq. 1}$$

where the first and second terms correspond to the contributions from nuclei and electron distribution, respectively. Alternatively, the i-th component of a dipole moment $\underline{\mu}$ can be calculated using the following formula

$$\underline{\mu_i} = -\frac{\partial E}{\partial \underline{\varepsilon_i}} \approx -\frac{\left(E\left(+\delta_{\underline{\varepsilon_i}}\right) - \left(E - \delta_{\underline{\varepsilon_i}}\right)\right)}{2\delta\varepsilon_i} \qquad \text{Eq. 2}$$

Equation 2 describes the response of the total energy (E) to a constant finite electric field $\underline{\varepsilon}$. This definition can be applied when the electronic densities are not available at a certain computational level (CCSD(T))

The dipole moment is one of the simplest quantities that is related to the distribution of electronic density in a polar molecule. In the absence of the dipole moment, the induction, long-range electrostatic interactions and the information incorporated in structure of infrared and sum-frequency generation spectra are uninterpretable.

In the gas phase the molecular dipole moment (μ) can be determined experimentally by microwave spectroscopy, molecular beam electric resonance spectroscopy, and other high-resolution spectroscopic techniques [1-8]. Experimentally measured values of μ in the liquid phase are very scarce and most often have large errors. Several indirect routes (X-ray diffraction experiment, Compton Scattering, dielectric spectra) exist for estimating the dipole moment of water, methanol..etc. molecules in condensed phases [5-8].

The most detailed studies on the value and variation of the dipole moment in different phases (vapour, liquid.. ) can be found for the water molecule. A dipole moment of 1.8546 Debye (D) in the gas phase was obtained [1,6]. Estimations of the experimental dipole moment in condensed phases fall between 2.6 and 2.9 D [2-5], where the lower limit (2.6 D) is related to a calculation based on the dielectric properties of hexagonal ice [4]. Indirect determinations of the dipole moment in the liquid state exist for acetone, acetonitrile and methanol molecules [2-5] . The dipole moment is increased by approximately 10 % compared to the gas phase for two non-hydrogen-bonded polar aprotic liquids (acetone and and acetonitrile). On the other hand, the dipole moment of methanol in the liquid state increases significantly, from approx. 1.7 to 2.87 D.

It is apparent that in condensed phases the dipole moment of polar molecules increases due to the interaction with the environment (polarization, charge transfer, geometrical deformation, etc. ...). At present, three methodically different approaches exist in the literature for calculating dipole moments of monomers in condensed phase. In the simplest case, one considers a single molecule described quantum mechanically (QM), using some reliable method and flexible basis set, that is embedded into an environment of point charges and/or multipoles determined by different assumptions. The coordinates of particles that build up the environment can be derived from simulations (Monte Carlo or Molecular Dynamics), or from local energy minimum configurations of various cluster sizes. According to this scheme, the dipole moment of water molecules increases by about 30 to 50% [9-11] in the liquid phase, while for acetone and acetonitrile the increase is estimated to be 10 to 15% [9]. These values depend on he applied QM method [9-11]. Even this scheme can account for the significant increase of the dipole moment as compared with a single monomer (in the gas phase).

The other two methods rely on the determination of a clear resolution of the total electron density for a given species. One of them can be applied at any level of theory, and and involves performing a three dimensional (3D) analysis based on Bader or Voronoi like decomposition scheme [12,13] into atomic (or, grouped together, to molecular) domains by studying the charge distribution. In this case the charge of the molecular domain can be different from 0.0.

Another approach for analysing the electronic properties of individual molecules in a supermolecular system (cluster) is by using a localisation technique to the molecule (or part of a molecule) in the Hilbert space. It is known that localized orbitals are not uniquely determined but depend on the localization criterion used. In this case, if we apply it to a neutral molecule, the net charge of the molecule is 0.0. Experience gained by reference [14] showed that it is practically irrelevant which type of localization (Boys [17], Natural localised Orbital [18] or two different version of the Magnasco-Perico [19,20] localization scheme) is used for water clusters. These methods results in dipole moment for water in clusters that are larger than those based on the Bader-type analysis by approximately 0.4-0.5 D [14,21,22].

Recent theoretical studies indicated a significant enhancement (up to ~2.9-3.05 D) of the dipole moment of individual water, methanol and acetonitrile molecules in clusters and in condensed phases [22-30]. It is conceivable that the more than 40 % increase of dipole moments of water and methanol molecules in the liquid state should be taken as a collective effect that is connected with the existence of H-bonded structures in these liquids. This collective effect can be well reproduced by considering relatively small water clusters, therefore such an approach helps in gaining deeper insight into its nature by performing various calculations on model water clusters. It has also been shown that the dipole moment of a water molecule depends on its hydrogen bonded local environment [31].

Liquid water and methanol are highly complex systems hence their theoretical and experimental studies continuously reveal new aspects of their structure [32-42]. The introduction of ions into bulk liquids causes a strong perturbation of the molecule structure in the close proximity of the cation. The various physical and chemical properties of ions in aqueous solutions are of evident importance in many biological, geological, and industrial processes [43-49]. Specific ionic effects are important in chemistry and biology. Such effects exhibit a recurring trend called the Hofmeister series [50-54]. In this series there are connections between effectiveness of salts in precipitating proteins and "their water ordering capacity".

There are different views concerning the importance of various effects in forming the modified water structure around an ion. In Refs. [55-66], it was assumed that collective effects between water molecules are the most crucial ones. However, in 1992 one of us [67] demonstrated computationally that the shortening of the distance between the first and second hydration shells around a cation is primarily caused by the direct electrostatic effect of the cation on the hydrogen bonds between water molecules: there is no need to assume any complicated collective effects to explain this experimentally (by diffraction) observed phenomenon.

Local polarization and charge transfer effects, induced by monovalent, divalent and trivalent cations in water, strongly influence structural and dynamical properties of the solvent (water). One of the manifestations of the polarisation terms is the perturbation of electronic density due to the ion-water interaction: this phenomenon can be characterised by the dipole moment of water molecules. Theoretical calculations revealed a significant increase of the dipole moment of water molecules around divalent ($Mg^{2+}$ 3.4 D [55,59], $Ca^{2+}$ 3.1 D [56,59,63]) and trivalent ($Al^{3+}$ 4.1 D [62,63]) cations. However, the dipole moment of water in the first shell of monovalent cations remains nearly the same (or slightly smaller than) as it is in bulk water [57-61].

Based on ab initio molecular dynamic simulation, Faralli et al. showed, by using ab initio molecular dynamics simulations, that the average dipole moment of methanol molecules in the first shell of $Li^+$ [66], $Ca^{2+}$ [64] and $Mg^{2+}$ [65] is significantly larger than in bulk liquid methanol. Also, in $Na^+$ and $K^+$ solutions the average value of dipole moment of methanol molecules in the first solvation shell is ∼ 0.2 D lower [66] than the average calculated for the entire volume of the solution.. These ions, in both the aqueous and methanolic solutions, have a similar effect on the dipole moment of the solvate molecules that build up solvation spheres.

In our recent paper [14] we showed that the dipole moment of a large water cluster can be reproduced only if one takes into account properly that there is a significant electron delocalization along the hydrogen bonds. The dipole moments of individual molecules contain significant components originating to the "tails" of the orbitals due to that delocalization, which even questions the concept of a water cluster consisting of individual molecules (as opposed to the concept of the whole clusters is considered as one single molecules). Building on the latter result, it was of interest to consider how the presence of a cation influences the dipole moments of the neighbouring solvate (water, methanol, formamide, acetone, acetonitrile) molecules. For this reason, we have compared results of three types of calculations. First, we calculated the dipole moments of the individual solvent molecules in clusters surrounding the cation. In the second type of calculations we kept the geometry of the clusters as obtained with ions, but replaced the latter with point-like charges of different values, then recalculated the wave functions of the resultant cluster. Next we examined how this change influences the values of the dipole moments. In the third type of calculations the ion was completely removed, but the arrangement of the molecules was kept the same as with ions. Again, the dipole moments were determined after recalculating of the cluster.

**Computational details and molecular orbital localisation method**

We have performed quantum chemical calculations for different cation-molecule clusters (first or first two hydration shells for water) with or without cations ($Li^+, Na^+, Mg^{2+}, Ca^{2+}, Al^{3+}$) by the Gaussian program suite Version 09 [67], using the the M05-2X DFT functional [68] and the cc-pVTZ basis set. We investigated three kinds of protic molecules, with the ability of acting as H-bond donors and acceptors (DA), methanol, formamide and water, and two aprotic polar molecules (A), acetone and acetonitrile. The coordination properties of the metal-DA moleculer complexes have been investigated in several articles [49,55-66,70-73]. In the M+6 DA molecule cluster (M=$Mg^{2+}, Ca^{2+}, Al^{3+}$) water, methanol and formamide molecules occupy octahedral positions around the metal ion, whereas these molecules in the inner sphere of $Li^+$ are accomodated on tetrahedral positions. Each water molecule of the inner sphere form two hydrogen bonds with outer shell waters molecules as H-bond donor. In the M+4A complexes solvent molecules occupy tetrahedral position.

In the $Na^+$+DA cluster, molecules form two virtually planar cyclic trimers (bonded through typical OH..O H-bonds or CH..O type H-bonds), above and below the ion.

The structures of these clusters are shown in Fig 1a,c,d for the first shell, and in Fig 1b for both the first shell and second shells of water. Atomic coordinates of these complexes can be found in Supp. Mat. Table S1. Additionally, the metal-ligand characteristic distances and complexation energies (with and without BSSE [74,75] correction and at aug-cc-pVTZ/cc-pVTZ level) are presented in the Supp. Mat. Table S2. At this level of theory (m052x/cc-pVTZ) the BSSE correction is about 1 to 9% (largest for $Na^+$--6methanol complex). The metal-ligand characteristic distances for $Li^+$, $Mg^{2+}$ and $Al^{3+}$ are similar (the difference is less than 5 %).

In order to calculate the dipole moment of a species in the clusters, the electronic density corresponding to the localized orbitals (for example, five and nine in the cases of water and methanol, respectively) of the given monomer needs to be investigated. It has been known that these orbitals are not uniquely determined and depend somewhat on the localization criterion applied [14]. The experience gained in reference [14] showed that it is practically irrelevant what localization scheme is used. Here, we use the scheme based on the Magnasco-Perico [19] localization criterion (with subsequent orthogonalization), by applying the algorithm developed by our group [15,16].

In these localization criteria the Mulliken net population of each orbital is maximized on a given individual monomer. Here we remark that if the delocalisation tail (contributions from different monomers, covalent bond order among the monomer) is too large than the exact interpretation of the monomer molecule becomes questionable. In these cases, we must interpret the obtained dipole moment values with great care. This localisation technique has the advantage that by introducing an additional procedure for truncating the orbitals to the basis of the defined monomer studied, one obtains an orthogonal set of orbitals. After normalizing these orbitals one can easily calculate the dipole moment of the monomer studied, without the influence of delocalization effects [14]. This way, a different localization transformation of the orbitals of the whole cluster is performed for each monomer separately. We denoted this type of technique as the Magnasco−Perico truncated (Mag-Per-trunc) scheme. In their previous work, Mayer et al. [14] found that for the dipole moment of water molecules in water cluster is in good agreement with that obtained by the Bader (3D type of decomposition scheme) method. (Further details are described in reference [14-16]. In that work we studied how the dipole moments of water molecules change when their number increases.).

For assessing the accuracy of our calculations, we have also calculated the dipole moments of the investigated monomers of the above mentioned, as well as of some other polar DA and A type molecules at the M052X/cc-pVTZ level of theory. There are several comprehensive benchmark data sets [76-80] for quantifying the the accuracy of different DFT functionals for calculating the dipole moment of molecules. In these works the authors mainly apply Eq. 2 at the CCSD(T)/aug-cc-pVQZ level as a benchmark value for calculation. In the present paper we also apply Eq 2 at the CCSD(T)/aug-cc-PVQZ (VTZ) level as a benchmark value. Additionally, we checked the accuracy of Eq. 2 at the above mentioned level of theory against Eq 1. It is possible to calculate the dipole moment of molecules at this level of theory since the electronic densities are available. In Eq. 2 we apply a field strength of 0.001 au (note that application of field strengths of an order of magnitude higher changes the results only insignificantly, by less than 0.3 %). The calculated data are presented in Table 1. These data clearly prove that the accuracy of Eq. 2 at the CCSD level of theory is better than 0.1 %. Additionally, these data also showed that the dipole moment of the molecules under study at the m052x/cc-pvTZ level of theory deviate from the benchmark theory by less than approximately 5 %. Thus the accuracy of our calculations appears sufficient to draw further conclusions.

Some of the localized orbitals belonging to water molecules near the ions are shown in Fig 2. In these molecular orbitals we can find significant contributions from atomic orbitals of the ions, as well ('back donation'). The actual charges on the metal ions, that can give us an insight into the extent of charge transfer process, were calculated using Bader [12,13], Mulliken and NBO [18] methods. The mixing term (contribution from different monomers in the MP scheme) or the Mayer bond order can provide information about the applicability of a given localisation scheme. If it is too large (Mayer bond order is close to 1 as a real covalent single bond) then the definition of a monomer, as well as calculation of the dipole moment itself, become questionable.

**Results and discussion**

*A: Dipole moment of water molecules around cations*

Table 2 contains data of some calculations devoted to assessing the qualitative and quantitative adequacy of the level of theory applied. For this purpose, we calculated the dipole moment for a single water molecule in the neighbourhood of an ion and/or point-like charge at the position of the ion. (The geometries correspond to ions and water molecules were taken from an optimised geometry of the ion+first hydration shell complex.) Point charge models are also considered for some charge values that are smaller than the nominal ones of the cations, in order to reflect the fact that effective cation charges are expected to be smaller than nominal ones.

Results obtained at the M05-2X/cc-pVTZ DFT level of theory are also compared with those calculated at the CCSD level —the highest level of theory for which dipole moment calculations are actually feasible without applying the approximation of Eq.2. CCSD calculations were performed for cases where the cation is represented by a point charge; the calculation of the dipole moment of water molecules separately would not be possible at the CCSD level if the ions were also treated at the full quantum mechanical level. For comparison, DFT and CCSD results obtained by using the aug-cc-pVTZ basis are also shown for the point charge models.

Inspecting data in Table 2, three main conclusions can be drawn. First, practically in all cases, the M05-2X DFT results are very close to those of the CCSD ones, giving us confidence in the adequacy of the DFT scheme applied. Second, the effect of a cation on the dipole moment is essentially the electrostatic effect of its charge. Third, the use of augmented basis set should be avoided in such calculations, as they lead to unrealistically large dipole moment values. Apparently, the presence of the diffuse functions lacking true localization on water leads to an overestimated relocation of the electron density to the positive center (overestimated "back-donation"). The value of dipole moment of water in M..L water complex for ($Li^+$, $Na^+$, $Mg^{2+}$, $Ca^{2+}$) ions also agrees well with results obtained from Krekeler et. al. [57-59], who used different functional and localisation scheme .

For reference, we quote some further numerical values. The experimental dipole moment of a single water molecule in vacuum is 1.855 D [1,6]; the dipole moment of a water molecule in bulk water is estimated (experimentally and theoretically) to be between 2.8 and 2.95 D [2-4,21-27]. Using the cc-pVTZ basis set and optimized geometry, the dipole moment of a free water molecule was obtained 1.986 D for the M05-2X DFT model and 1.908 D at the CCSD level of theory. Using the aug-cc-pVTZ basis these numbers reduce to 1.922 D and 1.850 D, respectively. The average value of the dipole moments of individual monomers in different clusters containing 20 to 30 water molecules was ca. 3.05 D at the M05-2X/cc-pVTZ level[31]. It is apparent from Table 2 that the presence of a cation (its positive charge) polarizes water molecule resulting in an increased dipole moment.

Tables 3 to 5 contain dipole moments of water molecules in water complexes containing the first two hydration shells around $Li^+$ and $Na^+$, $Mg^{2+}$ and $Ca^{2+}$, $Al^{3+}$, respectively. Reduced complexes, with the first hydration shell only, are also shown, along with results when the cations were replaced by point charges slightly smaller than their nominal values. These data indicate, again, that the dipole moments of water molecules are basically determined by the electrostatic charge of the cations. Values obtained for the cations are slightly lower than those obtained with the point charges equalling the formal charge of the cations, indicating that cations should be simulated using effective charges somewhat smaller than their nominal charges. This is in accordance with the common assumption that the presence of a bare positive charge induces some "back donation" of electrons from water molecules in the bulk. Note also that our results for dipole moments for water in the first shell of cations are in line with earlier studies, including more realistic sampling of the environment in a liquid phase (ab initio MD) [56-63].

Calculated charges on the metal atom in the clusters are shown in Table 7. Apparently, the calculated charges depend on the applied method (Bader, Mulliken, NBO, our method, i.e. mixing term in the Magnasco-Perico scheme), but all methods suggest some kind of charge transfer process. Our calculation based on the mixing of cation and water atomic orbitals give a well defined correlation (appr. 0.99) with the Mulliken charge. It is clear also that the influence of second shell water molecules to the cation's charges is not too large, but significant. This is not surprising since all of the second shell water molecules are H-bond acceptors. The effective charges, calculated using interpolation to recover the „true" dipole moment of water in the first shell of metals, are about 0.90 and 1.81 for mono and divalent cations, and about 2.72 for $Al^{3+}$. These data support the application of scaled charges for cations in molecular dynamics studies of aqueous solution [81-83].

Dipole moments of water molecules in the first hydration shell are, as expected, sensitive to the charge of the cation while, interestingly enough, the dipole moments in the second hydration shell are more-or-less the same for all complexes—although significantly larger than that of a free water molecule. This is not suprising since there are strong H-bonds between the first and second shells. The dipole moment of water in the water dimer is about 2.1-2.2 D, which is significantly smaller than the corresponding values in the second hydration shells. For ions with formal charges +2 and +3, the dipole moments of water molecules in the first hydration shell are significantly smaller than those observed for a single water molecule in the vicinity of a cation (Table 1). At the same time, the presence of the second hydration shell increases the dipole moments in the first shell, as may be observed by comparing results obtained for whole complexes with those in which only the first hydration shell is conserved. This fact is arising to the polarization effect of the water in the second shell.

The significant changes of dipole moments of water molecules in the vicinity of a cation with the increase of their number has already been indicated in Refs. [57-59]. This effect may be studied in some detail by considering our data in Table 6. These results indicate that—besides the electrostatic effect of the cation—cooperative effects of the water molecules play an important role in determining the value of the individual dipole moments. It is uncertain whether this effect is arising from the water molecules themselves in the octahedral (tetrahedral) arrangement (water-water interations) , or the polarised water molecules in the same position.

In order to obtain a better understanding of these collective effects, we have also performed calculations in which water molecules were considered in the molecular arrangement they have around a cation in the first hydration shell—but with the cations being removed. Results shown in Table 8 indicate that the geometrical arrangement of water molecules in the first hydration shell is such that their mutual polarization reduces the molecular dipole moments—in all cases, except the $Na^+$ complex, not only as compared to the bulk value but even below that of the free monomer. In the $Na^+$ complex the first hydration shell has a very specific hydrogen bonded network; this leads to some increase of the dipole moments due to the delocalizations along the hydrogen bonds among the water molecule. This counteracting behavior is essentially a manifestation of the Le Chatellier–Braun principle: the physico-chemical system reacts to the external perturbation in a manner so that to reduce the effect of the perturbation. In this case the perturbation to the dipole moment of water molecules is caused by the cation (electrostatically, quantum chemically), and water molecules occupy special positions around cations for reducing their dipole moment.

*B: Dipole moment of aprotic and protic molecules around a cation*

Calculated dipole moments of these molecules around cations are presented in Table 8. It is clear from the table that the dipole moments of all molecules are increased significantly, similarly to water complexes considered in more detail. We can detect the smallest/largest effects (ca . 30-50%/250-350%) for the cases of $Na^+/Al^{3+}$ for all solvent molecules, respectively. Here we would like to remark that the overlap term for $Al^{3+}$-solvent molecule complexes, and additionally, the $Al^{3+}$—O or N bond order is too large, so the calculated dipole moment for these complexes, using the MP procedure without truncation, can be questionable (Table 9 and Supp. Mat Table S3.)

These molecular dipole moments obtained for the cations are lower than those obtained with the point charges equal to the nominal charge of the cations, similarly to what we have already demonstrated for the case of M-water complexes. These results are presented in Fig 3.

Our results strongly suggest that cations should be simulated with an effective charge that is somewhat smaller (approx. 80 percent of the nominal charge) than the nominal charge. Here we would like to remark that the charges of ions in liquid water, applied in the scaled particle model, is also about 75-80 % of the nominal charge [81-83].

Additionally we showed that in the case of solvation of $Li^+$, $Mg^{2+}$ and $Al^{3+}$ ions, where the obtained energy minimum configurations give similar metal-ligand distances (see Supp. Mat Fig. S2a), the dipole moment of solvent molecules is approximately linearly proportional to the charge positioned into the location of the ion (Fig 3).

Distorting the geometry of solvent molecules does not cause a significant (greater than 2%) change in terms of their dipole moments, as presented in Supp. Mat Fig. S1. In these cases it was also observed that the dipole moment of the solvent molecules, without ions or charges in the centre of the complex, is significantly less than the value obtained for the optimal geometry. This finding is in good agreement with results obtained for ion water complexes (Table 8).

## Conclusions

From our calculations it can be concluded that the dipole moment of polar molecules in the neighborhood of a cation is primarily influenced by polarization from the bare electrostatic charge of the cation. The effective value of the ionic charge is somewhat reduced by "back donation" of electrons from water molecules. The effect of distorting molecular geometry on the dipole moment of solvent molecules is very small. A further important factor is the collective effect of water molecules: their mutual arrangement and polarization counteracts the polarizing effect of the cation, resulting in less increased values of the dipole moments, in comparison with the 1 ion – 1 solvent molecule setup. This behavior is essentially a manifestation of the Le Chatellier–Braun principle.

**Acknowledgment**

The authors acknowledge the financial support of the National Research, Development and Innovation Office (NRDIO (NKFIH), Hungary) via grants Nos 124885 and 128136. Computational resource from NIIF supercomputer center are acknowledged. Sz. Pothoczki acknowledges that this project was supported by the János Bolyai Research Scholarship of the Hungarian Academy of Sciences.

References

[1] W. S. Benedict, N. Gailar, E. K. Plyler, *J. Chem. Phys.* **1956**, *24*, 1139.

[2] Y. S. Badyal, M. -L. Saboungi, D. L. Price, S. D. Shastri, D. Haeffner, A. K. Soper, *J. Chem. Phys.* **2000**, *112*, 9206.

[3] J-K. Gregory, D.C.; Clary, K. Liu,M-G. ; Brown, R.J. Saykally, Science **1997**, 275, 814−817.

[4] C. A. D. Coulson, D. Eisenberg, Proc. R. Soc. London A **1966**, *291*, 445.

[5] D. H. Brookes, T. Head-Gordon, *J. Phys. Chem. Lett.* **2015**, *6,* 2938.

[6] S. A. Clough, Y. Beers, G. P. Klein, L. S. Rothman, *J. Chem. Phys.* **1973,** 59, 2254.

[7] O. Dorosh and Z. Kisiel, *Acta Physica Polonica* A **2007**, *112*, S-95.

[8] A. L. McClellan, Tables of Experimental Dipole Moment; Rahara Enterprises, El Cerrito, CA (1989).

[9] R. Rivelino, B. J. Costa Cabral, K. Coutinho, S. Canuto, *Chem. Phys. Lett.* **2005** *407*, 13.

[10] K. Coutinho, S. Canuto, *J. Chem. Phys.* **2000** *113*, 9132.

[11] S. Canuto, K. Coutinho, M. C. Zerner, *J. Chem. Phys.* **2000** *112*, 7293.

[12] R. F. W. Bader, Atoms in Molecules: A Quantum Theory; Oxford University Press: Oxford, U.K., 1990.

[13] R. F. W. Bader, C. F. Matta, *Int. J. Quantum Chem.* **2001**, *85*, 592.

[14] I. Bakó, and I. Mayer, *J. Phys. Chem. A* **2016**, *120*, 4408.

[15] I. Mayer, *Chem. Phys. Lett.* **1995**, *242*, 499.

[16] I. Mayer, *J. Phys. Chem.* **1996**, *100*, 6249.

[17] S. F. Boys, *Rev. Mod. Phys.* **1960**, *32*, 296.

[18] A. E. Reed, F. Weinhold, *J. Chem. Phys.* **1985**, 83, 1736.

[19] V. Magnasco, A. Perico, *J. Chem. Phys.* **1967**, 47, 971.

[20] I,Mayer, G.Räther, S.Suhai Chem. Phys. Letter 1998 293 81

[21] E. R. Batista, S. S. Xantheas, H. Jonsson, *J. Chem. Phys.* **2000**, *112*, 3285.

[22] L. D. Site, A. Alavi, R. M. Lynden-Bell, *Mol. Phys.* **1999**, *96*, 1683.

[23] A.V. Gubskaya, P. G. Kusalik, *J. Chem. Phys.* **2002**, *117*, 5290.

[24] P. L. Silvestrelli, M. Parrinello, *Phys. Rev. Lett.* **1999**, *82*, 3308.

[25] P.J.Dyer, P.; Cummings, P. J.Chem. Phys. **2006**, 125, 14451

[26] K. Laasonen, M. Sprik, M. Parrinello, R. Car, *J. Chem. Phys.* **1993**, *99*, 9080.

[27] P. L. Silvestrelli, and M: Parrinello, *J. Chem. Phys.* **1999**, *111*, 3572.

[28] R. F. Diasa, C. C. da Costaa, T. M. Manhabosco, A. B. de Oliveira, J. S. Matheus Matos, J. S. Soares, R. J. C. Batista, *Chem. Phys. Lett.* **2019**, *714*, 172.

[29] N. Sieffert, M. Bürl, M. Pierre Gaigeot, and C. A. Morrison, *J. Chem. Theory Comput.* **2013**, *9*, 106.

[30] J.-W. Handgraaf, T. S. van Erp, E. J. Meijer, *Chem. Phys. Lett.* **2003,** *367*, 617.

[31]. J. A. Morrone, K. E. Haslinger, M. E. Tuckerman J. Phys. Chem. B 2006, 110, 3712-3720

[32] I. Bakó, J. Daru, Sz. Pothoczki, L. Pusztai, K. Hermansson, *J. Mol. Liq.* **2019**, *293,* 111579.

[33] B. Bagchi, Water in Biological and Chemical Processes: From Structure and Dynamics to Function, Cambridge University Press, 2014.

[34] M. Chaplin, Water Structure and Science, London South Bank University http://www.lsbu.ac.uk/water/ (accessed 14 January, 2016).

[35] D. Eisenberg, W. Kauzmann, The Structure and Properties of Water, Oxford University Press: Oxford, 2005.

[36] F. Franks, Water, a Matrix of Life (2nd ed.), Royal Society of Chemistry: Cambridge, 2000.

[37] S. J. Grabowski, Hydrogen Bonding - New Insights, Ed. In Series Challenges and Advances in Computational Chemistry and Physics, Leszczynski, J., Ed.; Springer: New York, 2006.

[38] F. Sciortino, S. L. Fornili, *J. Chem. Phys.* **1989**, *90*, 2786.

[39] P. Kumar, G. Franzese, S. V. Buldyrev, H. E. Stanley, *Phys Rev E.* **2006**, *73*, 41505.

[40] I. Bakó, Á. Bencsura, K. Hermannson, Sz. Bálint, T. Grósz, V. Chihaia, J. Oláh, Phys Chem. Chem. Phys. **2013,** *15*, 15163.

[41] R. Lynden-Bell, S.C. Moris, J.D. Barrow, J. L. Finney, Jr. C. L. Harper, Water and Life: The Unique Properties of H2O, CRC Press: Boca Raton, 2010.

[42] P. Ball, *Chem. Rev.* **2008**, *108*, 74.

[43] L. X. Dang, T.-M. Chang, *J. Chem.Phys.* **2003**, *119*, 9851.

[44] H. Ohtaki, T. Radnai, *Chem. Rev.* **1993,** 92, 1157.

[45] H. J. Bakker, *Chem. Rev.* **2008**, *108*, 1456.

[46] Y. Marcus, *Chem. Rev.* **2009**, *109*, 1346.

[47] M. T. Ong, O. Verners, E. W. Draeger, A. C. T. Duin, V. Lordi, J. E. Pask, *J. Phys. Chem. B* **2015**, *119*, 1535.

[48] T. A. Pham, K. E. Kweon, A. Samanta, V. Lordi, J. E. Pask, *J. Phys. Chem. C* **2017,** *121*, 21913.

[49] R. López, N. Díaz, D. Suárez, *Chem. Phys. Chem.* **2020**, *21*, 99.

[50] L. Nostro Pierandrea, B.W. *Chem. Rev.* **2012**, *112*, 2286.

[51] N. Vlachy, B. Jagoda-Cwiklik, R. Vácha, D. Touraud, P. Jungwirth, W. Kunz, *Advances in Colloid and Interface Science* **2009**, *146*, 42.

[52] K. D. Collins, *Methods* **2004**, *34,* 300.

[53] J. Lyklema, *Chem. Phys. Lett.* **2009**, *467*, 217.

[54] Y. Zhang, P. S. Cremer, *Current Opinion in Chemical Biology* **2006**, *10*, 658.

[55] F. C. E. Lightstone, E. Schwegler, R. Q. Hood, F. Gygii, G. Galli, *Chem. Phys. Lett.* **2001,** *343*, 549.

[56] I. Bakó, J. Hutter, G. Pálinkas, *J. Chem. Phys.* **2002,** *117*, 9838.

[57] C. Krekeler, B. Hess, L. Delle Site, *J. Chem. Phys.* **2006**, *125*, 054305.

[58] C. Krekeler, L. Delle Site, *J. Phys.: Condens. Matter* **2007**, *19*, 192101.

[59] C. Krekeler, L. Delle Site, *J. Chem. Phys.* **2008**, *128*, 134515.

[60] T. Ikeda, M. Boero, K. Terakura, *J. Chem. Phys.* **2007**, *126*, 034501.

[61] Y. Liu, H. Lu, Y. Wu, T. Hu, Q. Li, *J. Chem. Phys*. **2010**, *132*, 124503.

[62] P. D. Mitev, I. Bakó, A. Eriksson, K. Hermansson, *Phys. Chem. Chem. Phys.* **2014,** *16*, 9351.

[63] S. Bogatko, E. Cauet, E. Bylaska, G. Schenter, J. Fulton, J. Weare, *Chem. Eur. J.* **2013,** *19*, 3047.

[64] C. Faralli, M. Pagliai, M. Cardini, V. Schettino, *J. Chem. Theory Comput.* **2008**, *4,* 156.

[65] M. Pagliai, G. Cardini, V. Schettino, *J. Phys. Chem. B* **2005**, *109*, 7475.

[66] E. Crabb, A. France-Lanord, G. Leverick, R. Stephens, Y. Shao-Horn, J. C. Grossman, *J. Chem. Theory Comput.* **2020**, *16*, 7255.

[67] I. Mayer, I. Lukovits, T. Radnai, *Chem. Phys. Lett.* **1992,** *188*, 595.

[68] M. Frisch, et al. Gaussian 09, Revision D.01, Gaussian, Inc., Wallingford CT, 2009.

[69] Y. Zhao, N. E. Schultz, D. G. Truhlar, *J. Chem. Theory Comput.* **2006**, *2*, 364.

[70] J. Songa, H. Kim, J. Seo, G. Yong, *J. Mol. Struct. THEOCHEM* **2004**, *686*, 147.

[71] M. Enrique, M. Cabaleiro-Lago, M. A. Rõos, *Chem. Phys.* **2000,** *254*, 11.

[72] A. J. Eilmes, P. Kubisiak, *J. Phys. Chem. A* **2010**, *114,* 973.

[73] G. Zhang, W. Wang, D. Chen, *Chem. Phys.* **2009,** *359*, 40.

[74] S. F. Boys, F. Bernardi, *Mol. Phys.* **1970**, *19*, 553.

[75] I. Mayer, I. Bako, *J. Chem. Theory Comput.* **2017**, *13*, 1883.

[76] D. Hait, M. Head-Gordon, *J. Chem. Theory Comput.* **2018**, *14*, 1969.

[77] A. L. Hickey, C. N. Rowley, *J. Phys. Chem. A* **2014**, *118*, 3678.

[78] A. S. Karne, N. Vaval, S. Pal, J. M. Vasquez-Perez, A. M. Köster, P. Calaminici, *Chem. Phys. Lett.* **2015**, *635*, 168.

[79] R. Grotjahn, G. J. Lauter, M. Haasler, M. Kaupp, *J. Phys. Chem. A* **2020**, *124*, 8346.

[80] M. Chołuj, J. Kozłowska, W. Bartkowiak, *Int. J Quantum Chem.* **2018**, e25666.

[81] M. Zeron, J. L. F. Abascal, C. A. Vega, *J. Chem. Phys.* **2019**, *151*, 134504.

[82] S. Blazquez, I.M. Zeron, M.M. Conde, J. L. F Abascal, C. Vega, *Fluid Phase Equilibria* **2020**, *513*, 112548.

[83] Z. R. Kann, J. L. Skinner, *J. Chem. Phys.* **2014** *141*, 104507.

Figure Captions:

Fig. 1. M..n-water investigated complexes: a: Octahedral and tetrahedral arrangements of water molecules around the metal cation, b: Arrangement of water molecules in the second shell of metal ion in M..(n-water+m-water, n=4,6 and m=8,12 respectively)c: arrangement of acetone andacetonitrile around metal cation, d: arrangement of formamide and methanol molecule around Na+ ion.

Fig. 2. Localised molecular orbitals of water around the metal ion that represent the most significant contributions from the empty orbitals of the ion

Fig. 3. The calculated dipole moment int he presence of charge/cation(cross) for the investigated cluster

Table 1. Dipole moment benchmarks for some simple polar protic and aprotic molecule. The molecular geometry was obtained using m052x/cc-pVTZ level of theory. The field strength applied in Eq. 2 was 0.001 au.

|  | ccsd/3 (Eq. 2) | ccsd(T) (Eq. 2) | ccsd/2 (Eq. 1) | ccsd/3 (Eq. 1) | m052x/a (Eq. 1) | m052x/3 (Eq. 1) |
|---|---|---|---|---|---|---|
| acetone | 3.0203 | 2.9522 | 3.0029 | 3.0010 | 3.1211 | 3.1996 |
| acetonitrile | 3.9741 | 3.9256 | 3.9482 | 3.9640 | 4.0979 | 4.1122 |
| methanol | 1.7228 | 1.6998 | 1.7174 | 1.7293 | 1.7213 | 1.7573 |
| ethanol | 1.6122 | 1.5882 | 1.6077 | 1.6180 | 1.6267 | 1.6419 |
| formic acid | 1.4795 | 1.4518 | 1.4724 | 1.4792 | 1.5763 | 1.5916 |
| acetic acid | 1.7279 | 1.6881 | 1.7178 | 1.7284 | 1.7979 | 1.8488 |
| formamide | 3.9238 | 3.9215 | 3.9110 | 3.9293 | 4.0807 | 4.091 |
| acetamide | 3.8865 | 3.8301 | 3.8728 | 3.8900 | 4.1048 | 4.032 |
| water | 1.8654 | 1.8365 | 1.8542 | 1.8645 | 1.9712 | 1.9591 |
| pyridine | 2.2691 | 2.2452 | 2.2676 | 2.2999 | 2.2563 | 2.3156 |

2: aug-cc-PVTZ, 3: aug-cc-pVQZ, a: cc-pVTZ

Table 2. Dipole moments of a single water molecule in the neighborhood of an ion and/or point-like charge (charge:ch)

| System | | | Dipole moment/D | | | | |
|---|---|---|---|---|---|---|---|
| Cation | Point charge | R(Me-O) Å | DFT+ion | DFT+p.c. | CCSD+p.c. | DFT-aug | CCSD-aug |
| Point ch. | 0.50 | 1.92 |  | 2.54 | 2.46 | 2.68 | 2.61 |
| Li+/point ch. | 1.00 | 1.92 | 3.13 | 3.15 | 3.07 | 3.66 | 3.59 |
| Point ch. | 0.50 | 2.38 |  | 2.25 | 2.26 | 2.31 | 2.24 |
| Na+/point ch. | 1.00 | 2.38 | 2.65 | 2.58 | 2.50 | 2.87 | 2.80 |
| Point ch. | 1.50 | 2.08 |  | 3.53 | 3.45 | 4.59 | 4.51 |
| Mg2+/point ch. | 2.00 | 2.08 | 4.28 | 4.11 | 4.02 | 6.06 | 6.13 |
| Point ch. | 1.50 | 2.37 |  | 3.21 | 3.12 | 3.94 | 3.12 |
| Ca2+/point ch. | 2.00 | 2.37 | 3.91 | 3.26 | 3.55 | 5.22 | 5.15 |
| Point ch. | 2.00 | 1.93 |  | 4.52 | 4.41 | 6.70 | 6.80 |
| Al3+/point ch. | 3.00 | 1.93 | 6.67 | 6.01 | 5.89 | 10.87 | 16.03 |

Table 3. Dipole moments of water molecules in the neighborhood of an ion and/or point-like charge calculated at the M05-2X/cc-pVTZ level for $Li^+$ and $Na^+$ complexes.

| | Dipole moment (D) | | | | |
|---|---|---|---|---|---|
| | point charge | 1st shell with cation | 2nd shell with cation | 1st shell with point ch. | 2nd shell with point ch. |
| point ch. with 12 $H_2O$ [a] | 0.75 | | | 3.03–3.07 | 2.69–2.72 |
| $Li^+$/point ch. with 12 $H_2O$ [a] | 1.0 | 3.11–3.13 | 2.61–2.63 | 3.20–3.25 | 2.71–2.75 |
| point charge with 4 $H_2O$ [b] | 0.75 | | | 2.56–2.57 | |
| $Li^+$/point ch. with 4 $H_2O$ [c] | 1.0 | 2.71-2.71 | | 2.83–2.84 | |
| point ch. with 18 $H_2O$ [d] | 0.75 | | | 3.02–3.05 | 2.69–2.83 |
| $Na^+$/point ch. with 18 $H_2O$ [d] | 1.0 | 3.08–3.11 | 2.61–2.63 | 3.10–3.14 | 2.72–2.83 |
| point charge with 6 $H_2O$ [e] | 0.75 | | | 2.48–2.50 | |
| $Na^+$/point ch. with 6 $H_2O$ [f] | 1.0 | 3.08–3.11 | | 2.61–2.63 | |

a: The geometry corresponds to the cluster of Li+ with 12 water molecules, b: The geometry corresponds to the cluster of Li+ with 4 water molecules, c: Only the first hydration shell of the previous cluster conserved, d: The geometry corresponds to the cluster of Na+ with 18 water molecules, e: The geometry corresponds to the cluster of Na+ with 6 water molecules, f: Only the first hydration shell of the previous cluster conserved

Table 4. Dipole moments of water molecules in the neighborhood of an ion and/or point-like charge calculated at the M05-2X/cc-pVTZ level for $Mg^{2+}$ and $Ca^{2+}$ complexes

| | Dipole moment (D) | | | | |
|---|---|---|---|---|---|
| | point charge | 1st shell with cation | 2nd shell with cation | 1st shell with point ch. | 2nd shell with point ch. |
| point ch. with 18 $H_2O$ [a] | 1.75 | | | 3.24-3.40 | 2.69-2.80 |
| $Mg^{2+}$/point ch. with 18 $H_2O$ [a] | 2.0 | 3.25-3.42 | 2.71-2.82 | 3.45-3.58 | 2.72-2.86 |
| point charge with 6 $H_2O$ [b] | 1.75 | | | 3.15-3.17 | |
| $Mg^{2+}$/point ch. with 6$H_2O$ [c] | 2.0 | 3.18-3.18 | | 3.39-3.41 | |
| point ch. with 18 $H_2O$ [d] | 1.75 | | | 3.13-3.35 | 2.71-2.84 |
| $Ca^{2+}$/point ch. with 18 $H_2O$ [d] | 2.0 | 3.19-3.40 | 2.73-2.90 | 3.33-3.49 | 2.73-2.89 |
| point charge with 6 $H_2O$ [e] | 1.75 | | | 3.02-3.05 | |
| $Ca^{2+}$/point ch. with 6 $H_2O$ [f] | 2.0 | 3.12-3.12 | | 3.22-3.24 | |

a: The geometry corresponds to the cluster of $Mg^{2+}$ with 18 water molecules, b: The geometry corresponds to the cluster of $Mg^{2+}$ with 6 water molecules, c: Only the first hydration shell of the previous cluster conserved, d: The geometry corresponds to the cluster of $Ca^{2+}$ with 18 water molecules, e: The geometry corresponds to the cluster of $Ca^{2+}$ with 6 water molecules, f: Only the first hydration shell of the previous cluster conserved

Table 5. Dipole moments of water molecules in the neighborhood of an ion and/or point-like charge calculated at the M05-2X/cc-pVTZ level for $Al^{3+}$ complexes

|  | Dipole moment (D) | | | | |
| --- | --- | --- | --- | --- | --- |
|  | point charge | 1st shell with cation | 2nd shell with cation | 1st shell with point ch. | 2nd shell with point ch. |
| point charge with 18 $H_2O$ [a] | 2.5 |  |  | 3.97-4.21 | 2.83-2.96 |
| $Al^{3+}$/point ch. with 18 $H_2O$ [a] | 3.0 | 4.09-4.26 | 2.99-3.12 | 4.61-4.84 | 2.98-3.12 |
| point charge with 6 $H_2O$ [b] | 2.5 |  |  | 4.06-4.09 |  |
| $Al^{3+}$/point Ch. with 6$H_2O$ [c] | 3.0 | 4.14-4.14 |  | 4.74-4.76 |  |

a: The geometry corresponds to the cluster of $Al^{3+}$ with 18 water molecules, b: The geometry corresponds to the cluster of $Al^{3+}$ with 6 water molecules, c: Only the first hydration shell of the previous cluster conserved, d: The geometry corresponds to the cluster of $Ga^{3+}$ with 18 water molecules, e: The geometry corresponds to the cluster of $Ga^{3+}$ with 6 water molecules, f: Only the first hydration shell of the previous cluster conserved

Table 6. Change of the dipole moments of water molecules (D) in the vicinity of a cation with the increase of their number

| Cation | Number of water molecule | | | | |
| --- | --- | --- | --- | --- | --- |
|  | 1 | 2 | 4[a] | 1st shell | 1st and 2nd shell |
| $Li^+$ | 3.13 | 2.97-2.98 | [b] | 2.72-2.72 | 3.11-3.13 |
| $Na^+$ | 2.65 |  | 2.68-2.69 | 2.56-2.56 | 3.08-3.11 |
| $Mg^{2+}$ | 4.28 | 4.08-4.08 | 3.52-3.57 | 3.18-3.18 | 3.25-3.42 |
| $Ca^{2+}$ | 3.91 | 3.70-3.71 | 3.34-3.37 | 3.12-3.12 | 3.19-3.40 |
| $Al^{3+}$ | 6.67 | 6.01-6.01 | 4.75-4.85 | 4.14-4.14 | 4.08-4.20 |

a: in the case of $Na^+$   b: The first hydration shell contains only 4 water molecules

Table 7. The calculated charge on metal ion the the M-(Wa)$_n$ cluster

|  | Bader | Mulliken | NBO | mixing term |
| --- | --- | --- | --- | --- |
| $Li^+$+4$H_2O$ | 0.912 | 0.191 | 0.720 | 0.065 |
| $Li^+$+4$H_2O$+8$H_2O$ | 0.901 | 0.010 | 0.690 | 0.010 |
| $Na^+$+6$H_2O$ | 0.913 | 0.353 | 0.760 | 0.250 |
| $Na^+$+6$H_2O$+12$H_2O$ | 0.904 | 0.263 | 0.680 | 0.097 |
| $Mg^{2+}$+6H2O | 1.790 | 0.822 | 1.381 | 0.732 |
| $Mg^{2+}$+6$H_2O$+12$H_2O$ | 1.781 | 0.681 | 1.342 | 0.625 |
| $Ca^{2+}$+6$H_2O$ | 1.732 | 1.032 | 1.441 | 0.961 |
| $Ca^{2+}$+6$H_2O$+12$H_2O$ | 1.713 | 0.781 | 1.334 | 0.718 |
| $Al^{3+}$+6$H_2O$ | 2.642 | 0.471 | 1.981 | 0.348 |
| $Al^{3+}$+6$H_2O$+12$H_2O$ | 2.631 | 0.323 | 1.974 | 0.174 |

Table 8. Dipole moments of polar and aprotic molecules of the first hydration shell with the cation removed. Dipole moment are given in Debye (D), Dmon: dipole moment of the monomer in optimized geometry

|    | Methanol mon:1,72D | Formamide mon=4,08D | Acetonitrile mon:4,09 | Acetone mon:3,12 | Water mon:1,98D |
|----|---|---|---|---|---|
| Li | 1,43 | 3,91 | 3,28 | 2,69 | 1,77 |
| Na | 1,89 | 4,02 | 3,57 | 2,85 | 2,15 |
| Mg | 1,38 | 3,73 | 3,39 | 2,69 | 1,61 |
| Ca | 1,43 | 3,74 | 3,56 | 2,79 | 1,75 |
| Al | 1,32 | 3,69 | 3,28 | 3,27 | 1,52 |

Table 9. The dipole moment of molecule in the M-nL (n=4,6) complexes using Magnasco-Perico [14,19] (MP-1) or truncated Magnasco-Perico [14] (MP-2) localisation technique. (Dmon: dipole moment of the monomer in optimized geometry)

| Cation | Methanol | (Dmon:1.7213D) | Formamide | (Dmon=4.0807D) |
|---|---|---|---|---|
| n=4 | MP-1 | MP-2 | MP-1 | MP-2 |
| $Li^+$ | 2.254 | 2.891 | 5.273 | 5.852 |
| $Na^+$ | 2.284 | 2.634 | 4.951 | 5.334 |
| $Mg^{2+}$ | 2.901 | 3.681 | 6.067 | 6.836 |
| $Ca^{2+}$ | 2.917 | 3.591 | 4.952 | 5.321 |
| $Al^{3+}$ | 3.551 | 5.182 | 7.309 | 8.759 |
|  | Methanol | (Dmon:1.7213D) | Formamide | (Dmon=4.0807D) |
| n=6 | MP-1 | MP-2 | MP-1 | MP-2 |
| $Li^+$ | 2.84 | 3.01 | 5.76 | 5.95 |
| $Na^+$ | 2.48 | 2.66 | 5.18 | 5.34 |
| $Mg^{2+}$ | 3.63 | 3.89 | 6.78 | 7.04 |
| $Ca^{2+}$ | 3.47 | 3.67 | 6.58 | 6.76 |
| $Al^{3+}$ | 5.02 | 5.83 | 8.67 | 9.41 |

Fig. 1. M..n-water investigated complexes: a: Octahedral and tetrahedral arrangements of water molecules around the metal cation, b: Arrangement of water molecules in the second shell of metal ion in M..(n-water+m-water, n=4,6 and m=8,12 respectively)c: arrangement of acetone andacetonitrile around metal cation, d: arrangement of formamide and methanol molecule around Na+ ion.

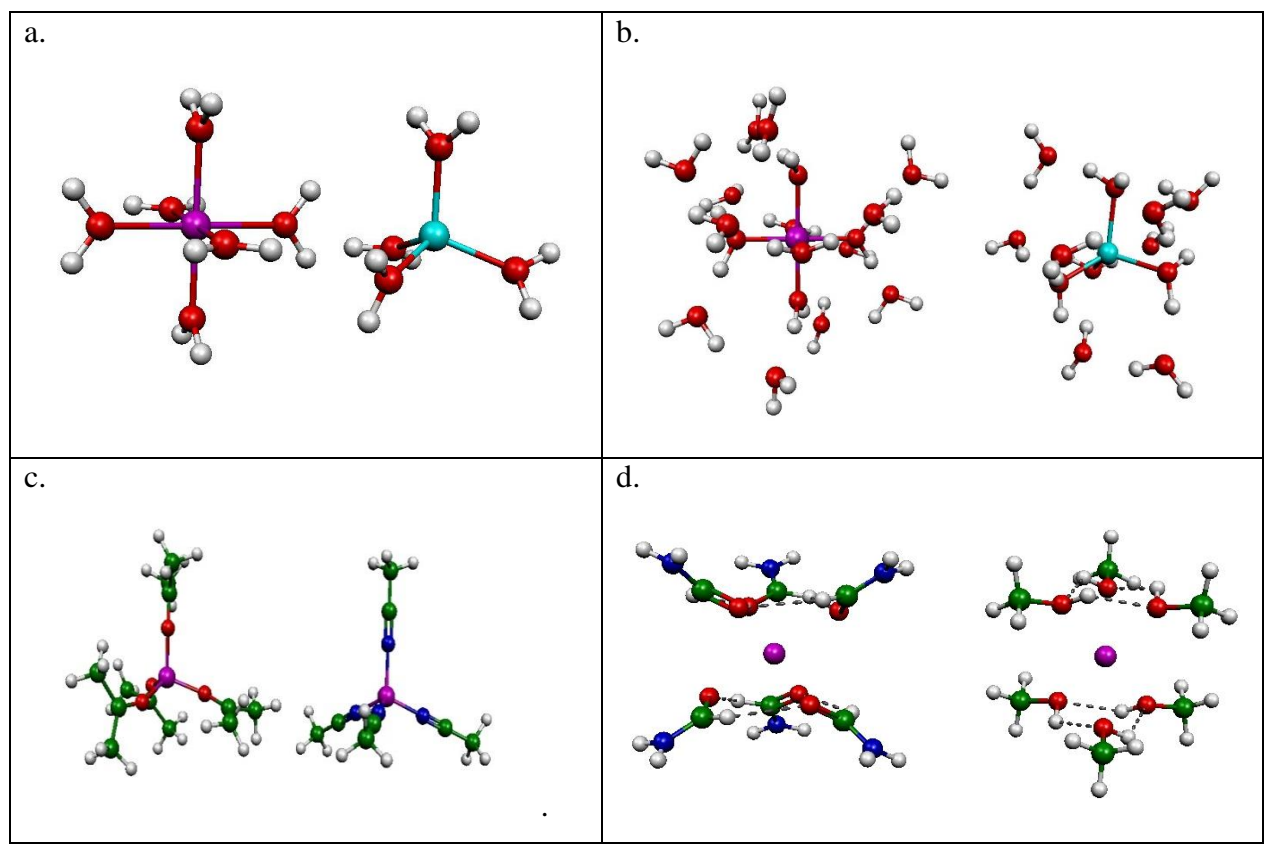

Fig. 2. Localised molecular orbitals of water around the metal ion that represent the most significant contributions from the empty orbitals of the ion.

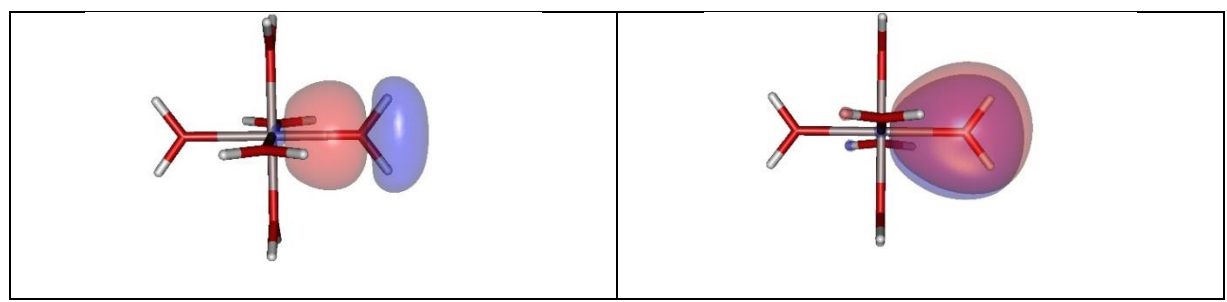

Figure 3. The calculated dipole moment int he presence of charge/cation(cross) for the investigated cluster

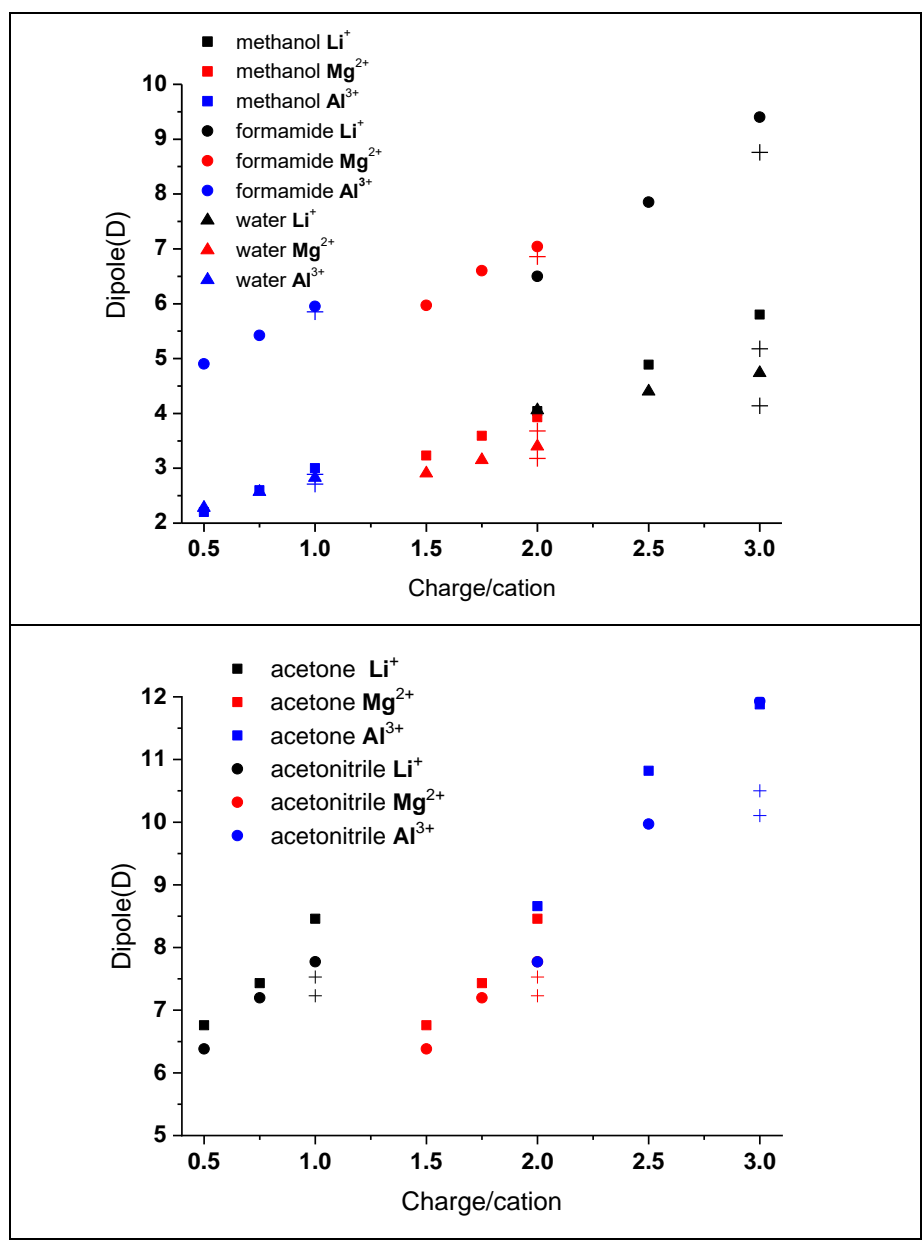

Supporting Information

for

# Influence of cations on the dipole moments of neighboring polar molecules

Imre Bakó, Dániel Csókás and István Mayer
Institute of Organic Chemistry, Research Centre for Natural Sciences,
H-1519 Budapest, P.O.Box 286, Hungary

Szilvia Pothoczki[1], László Pusztai[1,2]
[1]Wigner Research Centre for Physics, H-1121 Budapest, Konkoly Thege M. út 29-33.,
Hungary
[2]International Research Organization for Advanced Science and Technology (IROAST),
Kumamoto University, 2-39-1 Kurokami, Chuo-ku, Kumamoto, 860-8555, Japan

Table S1. Coordinate of M-nL (L: water,methanol, formamide, acetone, acetonitrile, n:4 or 6 or 12 or 18) cluster in optimised geometry at m052x/cc-pVTZ level of theory

**Li+.. 4wa**
**13**
Li 0.007825 0.006481 0.005372
 O -1.251213 -1.093696 0.961694
 H -1.194358 -1.297648 1.897623
 H -2.080413 -1.460751 0.648842
 O -0.957865 1.296279 -1.047214
 H -1.729355 1.811464 -0.801635
 H -0.782829 1.480498 -1.971708
 O 1.157788 0.909294 1.260841
 H 1.903826 0.543767 1.740990
 H 1.138421 1.848468 1.453688
 O 1.050099 -1.101227 -1.184305
 H 0.844865 -2.016948 -1.378152
 H 1.896111 -0.916393 -1.596295

**Li+.. 4wa..8wa (first and second shell)**
**37**
Li 0.000449 -0.011634 0.026974
 O -1.346327 -0.448930 1.367863
 H -0.987280 -0.755415 2.232889
 H -1.913320 -1.135301 0.988877
 O -1.226019 0.588107 -1.364076
 H -1.708881 1.347140 -1.007913
 H -0.791889 0.839355 -2.212433
 O 1.364653 1.267351 0.580212
 H 1.849157 0.864690 1.314592
 H 1.020538 2.149369 0.853641
 O 1.209032 -1.458877 -0.467093
 H 0.760692 -2.287315 -0.756614
 H 1.774090 -1.128121 -1.179413
 O -2.352651 2.080485 0.758392
 H -2.172524 1.232571 1.201649
 H -3.241680 2.349507 0.997383
 O 0.306172 0.979077 -3.479077
 H 0.305678 1.543238 -4.252905
 H 1.207579 0.953897 -3.123232
 O 2.564081 0.603384 -1.848915
 H 3.483156 0.801377 -2.038466
 H 2.329964 1.056848 -1.019783

O -0.376018 -3.505302 -0.993409  
H -0.363412 -4.278577 -1.558151  
H -1.260848 -3.111867 -1.038421  
O 0.023778 -1.037634 3.547524  
H -0.081984 -1.598450 4.316506  
H 0.935654 -1.126338 3.230792  
O -2.584911 -1.779453 -0.797757  
H -3.492054 -1.928701 -1.070977  
H -2.276589 -0.964532 -1.232190  
O 0.050215 3.510116 1.044098  
H 0.141805 4.275780 1.611918  
H -0.878993 3.234175 1.054872  
O 2.376863 -0.950677 2.014018  
H 3.254513 -1.264174 2.240509  
H 2.121202 -1.372236 1.174498  

**Na$^+$.. 6wa**
**19**
Na 0.000000 0.000000 0.000000  
 O -2.058600 -0.034840 1.206630  
 H -1.973480 -0.980150 1.383930  
 H -2.505250 0.364970 1.954500  
 O -1.828910 -0.373630 -1.486450  
 H -2.464710 -0.052520 -0.834610  
 H -2.157320 -0.140740 -2.356260  
 O -0.800650 -2.237390 0.213020  
 H -0.502360 -3.140790 0.329580  
 H -1.263540 -2.185690 -0.632750  
 O 2.057890 0.034520 -1.206750  
 H 1.972940 0.979660 -1.384930  
 H 2.504820 -0.366040 -1.954060  
 O 1.829040 0.372650 1.486480  
 H 2.157380 0.139170 2.356170  
 H 2.464760 0.051710 0.834480  
 O 0.800870 2.237350 -0.212720  
 H 1.264460 2.185680 0.632660  
 H 0.501990 3.140640 -0.328660  

**Na$^+$..6wa..12 wa**
**55**
Na -0.008170 0.004415 0.028315  
 O -0.396282 -2.133744 -1.081136  
 H -0.377212 -2.894568 -0.459990  
 H -1.284899 -2.095638 -1.467350  
 O 1.599081 0.612572 -1.750409  
 H 1.748955 -0.225000 -2.222874  
 H 2.460043 0.869001 -1.373880

O 1.708114 -0.952201 1.421873
H 1.328321 -1.591876 2.042863
H 2.353468 -1.413451 0.832345
O 0.381823 2.143633 1.137006
H 1.270029 2.105529 1.524081
H 0.363459 2.903765 0.515026
O -1.617605 -0.603196 1.804805
H -2.479527 -0.857995 1.429520
H -1.765189 0.234771 2.277336
O -1.725114 0.961547 -1.364659
H -2.370455 1.423301 -0.775477
H -1.345757 1.600724 -1.986452
O 4.038777 1.049376 -0.460261
H 3.814892 1.282077 0.460222
H 4.757209 1.621194 -0.736131
O 1.805711 -2.014090 -2.608169
H 0.950260 -2.211645 -2.166370
H 1.846736 -2.508980 -3.427386
O -4.059401 -1.036463 0.516503
H -4.778708 -1.607447 0.791811
H -3.833881 -1.270578 -0.403200
O -1.820777 2.023704 2.662976
H -0.965372 2.221745 2.221177
H -1.861754 2.517405 3.482906
O 0.298669 4.125704 -0.683099
H 0.185830 3.754255 -1.572840
H -0.217626 4.931087 -0.637922
O 2.922868 1.324135 2.050641
H 2.530346 0.419722 2.010106
H 3.377694 1.416831 2.889323
O -0.312260 -4.117220 0.737113
H 0.205606 -4.921647 0.692772
H -0.201440 -3.745661 1.627033
O -2.939047 -1.315066 -1.991905
H -2.546447 -0.410623 -1.951724
H -3.392344 -1.408623 -2.831318
O -0.066920 -2.482979 2.970573
H -0.739644 -1.823334 2.694888
H -0.190812 -2.664735 3.903183
O 3.512655 -1.813496 -0.332652
H 3.099992 -2.048053 -1.176152
H 3.986744 -0.986244 -0.485730
O -3.529539 1.825639 0.388649
H -4.005160 0.999183 0.541390
H -3.116264 2.059021 1.232193
O 0.047625 2.491393 -2.916142
H 0.720321 1.831467 -2.640853

H 0.170440 2.672646 -3.848988

**Mg$^{2+}$..6wa**
**19**
Mg 0.000000 0.000000 0.000000
O -0.855480 -1.686060 -0.882660
H -0.658580 -2.605270 -0.668280
H -1.521940 -1.695040 -1.579330
O 1.510440 -0.013000 -1.440370
H 1.607510 -0.643180 -2.163640
H 2.242740 0.610490 -1.508670
O 1.157900 -1.230390 1.225170
H 0.915800 -1.561530 2.097730
H 2.036000 -1.575110 1.025830
O 0.857220 1.684540 0.884750
H 1.523760 1.687610 1.581420
H 0.664320 2.605520 0.674450
O -1.507850 0.010560 1.443040
H -2.240740 -0.612240 1.511340
H -1.604830 0.641340 2.165790
O -1.157060 1.229520 -1.224490
H -2.036020 1.572740 -1.026310
H -0.914190 1.561920 -2.096350

**Mg$^{2+}$..6wa..12 wa**
**55**
Mg 0.038650 0.042439 0.332186
O -1.147623 -1.466859 -0.536956
H -0.630838 -2.307788 -0.600396
H -1.507034 -1.314595 -1.433183
O 1.545467 -0.325312 -1.099617
H 1.296027 -0.764151 -1.936673
H 1.971754 0.526264 -1.366150
O 0.846319 -1.365248 1.585408
H 0.702926 -1.405935 2.546196
H 1.630572 -1.885106 1.349460
O 1.231792 1.491275 1.157746
H 1.645117 1.436732 2.036087
H 1.213644 2.414832 0.861645
O -1.433264 0.448650 1.704725
H -2.281796 -0.021267 1.711813
H -1.293929 0.885178 2.562180
O -0.834319 1.459928 -0.964999
H -1.820472 1.415813 -0.902671
H -0.631086 1.404912 -1.919477
O 2.381729 2.043105 -1.995988
H 1.998937 2.802365 -1.531081

```
H 3.255461 2.302532 -2.295648
O 0.821526 -1.298143 -3.596234
H -0.066935 -1.685046 -3.608266
H 1.389266 -1.863467 -4.126020
O -3.563907 -1.093172 0.901635
H -4.254786 -1.659291 1.253912
H -2.904958 -1.650497 0.463901
O -0.751408 1.387632 4.245406
H 0.149002 1.746781 4.250173
H -1.283399 1.928912 4.834456
O 0.775300 3.722195 -0.379592
H -0.068763 3.305736 -0.604653
H 0.636338 4.669238 -0.304212
O 2.024926 1.274818 3.828462
H 1.921358 0.365834 4.149047
H 2.815626 1.638099 4.235663
O 0.435395 -3.567164 -0.972874
H 0.221014 -4.502099 -0.952741
H 1.333630 -3.469081 -0.622678
O -1.939513 -1.187056 -3.183532
H -1.865770 -0.277032 -3.508693
H -2.761916 -1.548417 -3.523650
O 0.585703 -1.081599 4.351123
H -0.127007 -0.462208 4.570573
H 0.581489 -1.776373 5.014634
O 2.842246 -2.541318 0.107016
H 3.745642 -2.856843 0.185867
H 2.847866 -1.721560 -0.407193
O -3.464472 1.026756 -0.970358
H -3.757142 0.355318 -0.335762
H -4.180439 1.657667 -1.069663
O -0.511321 1.156054 -3.705257
H 0.219252 0.562263 -3.935192
H -0.483816 1.896591 -4.316345
```

**Ca$^{2+}$..6wa**
**19**
```
Ca 0.000000 0.000000 0.000000
 O -0.215340 2.064740 1.142300
 H 0.381850 2.427720 1.805240
 H -0.917890 2.715640 1.039110
 O -1.245520 -1.074420 1.705730
 H -1.620490 -0.668880 2.494930
 H -1.482670 -2.006840 1.753170
 O 2.004600 -0.445590 1.182870
 H 2.899880 -0.205400 0.920820
 H 2.092910 -0.902710 2.026160
```

O 0.215630 -2.064420 -1.142840
H -0.381620 -2.427260 -1.805800
H 0.918210 -2.715330 -1.039880
O 1.245250 1.074710 -1.705680
H 1.482360 2.007150 -1.752800
H 1.620180 0.669460 -2.495040
O -2.005100 0.445260 -1.182140
H -2.094070 0.902330 -2.025370
H -2.900130 0.204710 -0.919560

**Ca$^{2+}$..6wa..12wa**
**55**
Ca 0.051477 0.049444 0.401478
 O -1.316796 -1.608193 -0.680548
 H -0.803565 -2.451514 -0.741576
 H -1.637925 -1.430699 -1.585426
 O 1.700404 -0.393646 -1.294382
 H 1.407835 -0.789924 -2.137547
 H 2.137071 0.459103 -1.539413
 O 1.100826 -1.504500 1.788412
 H 1.086208 -1.527365 2.760194
 H 1.817280 -2.067401 1.457756
 O 1.290193 1.765068 1.382809
 H 1.633121 1.797125 2.291901
 H 1.310876 2.652106 0.992685
 O -1.620085 0.347167 2.000798
 H -2.487353 -0.074296 1.902675
 H -1.509168 0.662817 2.913580
 O -0.904331 1.645618 -1.123023
 H -1.889506 1.607012 -1.044749
 H -0.714052 1.553688 -2.076416
 O 2.550152 2.025430 -2.007254
 H 2.129874 2.771936 -1.552619
 H 3.431167 2.305430 -2.263589
 O 0.787582 -1.269712 -3.775423
 H -0.106258 -1.645005 -3.766857
 H 1.319932 -1.806042 -4.368522
 O -3.682116 -1.028657 0.772794
 H -4.424798 -1.585844 1.017664
 H -3.025922 -1.581791 0.321859
 O -0.845676 1.056778 4.595848
 H -0.037020 1.591491 4.561910
 H -1.412733 1.437399 5.271667
 O 0.886507 3.754003 -0.496007
 H 0.039678 3.343586 -0.728827
 H 0.774436 4.707202 -0.524837
 O 1.886700 1.564909 4.110087

H 1.973827 0.631076 4.357876
H 2.570415 2.050803 4.578706
O 0.312371 -3.686938 -0.995570
H 0.112643 -4.623392 -0.936601
H 1.219597 -3.563687 -0.675584
O -1.983640 -1.180082 -3.350831
H -1.898919 -0.259757 -3.643496
H -2.777267 -1.539347 -3.755677
O 0.999453 -1.076699 4.553053
H 0.172650 -0.625444 4.784986
H 1.168622 -1.732802 5.234280
O 2.792269 -2.696309 -0.046276
H 3.675509 -3.071835 -0.075519
H 2.797976 -1.872312 -0.556899
O -3.522582 1.197949 -0.973547
H -3.795216 0.490264 -0.368943
H -4.243741 1.828970 -1.019253
O -0.563271 1.188471 -3.847446
H 0.163579 0.585814 -4.067217
H -0.560231 1.893822 -4.499583

**Al3+..6wa**
**19**
Al 0.000000 0.000000 0.000000
O -0.792640 -1.552810 -0.821560
H -0.595360 -2.482580 -0.608880
H -1.467790 -1.553860 -1.523550
O 1.390520 -0.005020 -1.334580
H 1.484530 -0.639060 -2.067920
H 2.130040 0.625340 -1.400830
O 1.073370 -1.141390 1.122860
H 0.830270 -1.479910 2.003150
H 1.959410 -1.487970 0.914480
O 0.792580 1.552790 0.821640
H 1.468110 1.553820 1.523260
H 0.595040 2.482570 0.609230
O -1.390560 0.005020 1.334550
H -2.130650 -0.624710 1.400170
H -1.483910 0.638350 2.068590
O -1.073350 1.141400 -1.122840
H -1.959260 1.488200 -0.914310
H -0.830420 1.479650 -2.003290

**Al$^{3+}$+6wa+12wa**
**55**
Al -0.026229 -0.012120 -0.095187
O -0.898072 -1.497860 -0.912968

```
H -0.884210 -2.415970 -0.567722
H -1.395586 -1.457921 -1.764913
O 1.335097 -0.120874 -1.426770
H 1.256484 -0.628877 -2.269557
H 2.195048 0.349680 -1.389790
O 0.939162 -1.211233 1.010123
H 0.736653 -1.281038 1.973857
H 1.941303 -1.251194 0.909949
O 0.868368 1.474020 0.668398
H 1.335257 1.392541 1.534355
H 0.404925 2.369053 0.657877
O -1.415062 0.114125 1.187730
H -1.931274 -0.720746 1.417005
H -1.283750 0.619679 2.025362
O -1.007676 1.152321 -1.244002
H -1.783837 1.686525 -0.971128
H -0.829979 1.270853 -2.207899
O 3.896269 0.878557 -1.219518
H 4.092166 1.787466 -0.967851
H 4.543519 0.657251 -1.898999
O 0.902821 -1.204933 -3.843189
H 0.069309 -1.690597 -3.925572
H 1.542837 -1.631981 -4.422417
O -2.512748 -2.094426 1.908835
H -3.372640 -2.242676 2.312058
H -2.227610 -2.921594 1.497704
O -0.942009 1.209675 3.600686
H -0.103080 1.684891 3.687606
H -1.584120 1.649245 4.167347
O -0.471286 3.660839 0.833408
H -1.375143 3.691088 0.492027
H -0.154974 4.564226 0.921488
O 1.854668 1.334239 3.169281
H 1.891981 0.446907 3.554439
H 2.599890 1.833240 3.518993
O -1.197748 -4.010655 0.181646
H -1.763029 -4.653572 -0.261873
H -0.465460 -4.521877 0.542537
O -1.899654 -1.394892 -3.398299
H -1.943307 -0.506529 -3.780843
H -2.636156 -1.899795 -3.759025
O 0.610323 -1.159453 3.681326
H -0.145016 -0.644633 4.000008
H 0.734967 -1.896416 4.287946
O 3.502307 -1.073055 0.883988
H 4.159108 -1.730211 1.130495
H 3.913722 -0.456036 0.263822
```

O -2.956466 2.942031 -0.468942
H -3.707621 2.702757 0.084525
H -3.305209 3.566893 -1.115083
O -0.708942 1.134714 -3.909718
H 0.051524 0.626617 -4.227528
H -0.845163 1.863369 -4.524516

25
**Li+ 4 methanol**
Li -0.183664 -0.194203 0.059255
O -1.431173 -0.784713 1.397820
H -1.524469 -0.370303 2.257175
C -2.314310 -1.917175 1.312822
H -2.081326 -2.638376 2.090006
H -3.348666 -1.599060 1.396891
H -2.149378 -2.366582 0.341209
O -1.040043 0.881967 -1.291569
H -1.989474 0.919586 -1.419659
C -0.390840 1.767073 -2.220184
H -0.719077 2.790187 -2.063481
H 0.672531 1.694513 -2.025311
H -0.595976 1.463960 -3.242442
O 1.162035 0.993757 0.759703
H 2.050564 0.723218 0.997093
C 0.981026 2.384395 1.078244
H 1.109922 2.550933 2.143118
H 1.681377 2.996333 0.517901
H -0.032142 2.638650 0.790932
O 0.555461 -1.793345 -0.722631
H 0.889655 -2.529608 -0.207389
C 0.805631 -2.034977 -2.118963
H 1.872946 -2.065767 -2.314199
H 0.340481 -2.964688 -2.430673
H 0.358672 -1.210797 -2.660663

25
**Li+ 4 formamide**
Li 0.004364 -0.002653 -0.000976
C 0.332209 2.639888 0.872505
N 0.075350 3.755140 1.549134
O -0.450090 1.698563 0.789786
H 1.305328 2.613625 0.378848
H 0.745908 4.497777 1.597429
H -0.793652 3.843639 2.045417
C -0.306417 -2.510952 1.209594
N -0.037572 -3.530716 2.019162
O 0.474100 -1.589631 0.993075

H -0.707085 -4.259569 2.174149
H 0.840162 -3.556765 2.507208
H -1.288197 -2.546092 0.733924
C -2.585391 0.249995 -1.042065
N -3.661155 -0.059564 -1.759345
O -1.650287 -0.520474 -0.850219
H -4.399531 0.603047 -1.898006
H -3.722175 -0.962750 -2.195010
H -2.586460 1.256673 -0.620183
C 2.576541 -0.389430 -1.043197
N 3.639952 -0.176118 -1.812095
O 1.643662 0.400168 -0.937194
H 4.377305 -0.851264 -1.875315
H 3.692398 0.662904 -2.362150
H 2.586265 -1.332586 -0.493924

41
**Li+ 4 acetone**
Li -0.057966 0.108717 -0.117810
O -1.528419 -0.942961 0.548429
C -1.913219 -1.619958 1.485289
C -3.255837 -2.275791 1.460772
C -1.067453 -1.830094 2.701209
H -0.075412 -1.420682 2.550528
H -1.546929 -1.336633 3.546427
H -1.018578 -2.889860 2.942708
H -3.792839 -2.065162 2.383201
H -3.824671 -1.944579 0.600884
H -3.115642 -3.355598 1.415875
O -0.833418 1.731838 -0.826247
C -1.679006 1.930214 -1.682910
C -2.294561 0.804824 -2.450744
C -2.145834 3.313649 -1.996458
H -1.785506 4.016719 -1.255674
H -1.763956 3.588840 -2.979821
H -3.231415 3.339464 -2.060729
H -2.447574 1.086200 -3.489317
H -1.687781 -0.089601 -2.372882
H -3.277495 0.609823 -2.020019
O 1.247681 0.481095 1.268494
C 1.815894 1.414990 1.812194
C 2.755895 1.180692 2.949875
C 1.605745 2.827240 1.373750
H 0.779958 2.894259 0.675896
H 1.442273 3.466774 2.238453
H 2.522463 3.173884 0.895645
H 3.677452 1.738358 2.798535

H 2.298611 1.568977 3.860076
H 2.960163 0.123332 3.066956
O 0.739040 -0.883657 -1.564466
C 1.652407 -1.645072 -1.830525
C 2.597599 -2.132548 -0.778680
C 1.869700 -2.128496 -3.227352
H 2.785735 -1.677908 -3.609518
H 2.019896 -3.205772 -3.235173
H 1.035810 -1.850754 -3.859976
H 2.423202 -3.197627 -0.626072
H 3.625147 -2.028234 -1.120481
H 2.443292 -1.596176 0.150398

37
**Na+ 6methanol**
Na -0.000012 -0.000969 -0.001282
O -2.039880 -0.029640 1.263867
H -1.965934 -0.981398 1.397078
C -2.718692 0.579440 2.368953
H -3.724539 0.181972 2.468864
H -2.168219 0.424074 3.292779
H -2.774160 1.640764 2.159637
O -1.836228 -0.325146 -1.510207
H -2.466311 -0.047198 -0.835851
C -2.320925 0.008757 -2.816587
H -3.259825 -0.498065 -3.019674
H -2.453481 1.082615 -2.915883
H -1.574181 -0.326708 -3.525912
O -0.760479 -2.271671 0.185577
H -1.257619 -2.193016 -0.636456
C -0.323273 -3.622468 0.378634
H 0.332293 -3.934815 -0.429607
H 0.224010 -3.647021 1.313063
H -1.172805 -4.296021 0.444336
O 2.039673 0.027668 -1.266778
H 1.965839 0.979447 -1.399893
C 2.718154 -0.581418 -2.372063
H 3.724064 -0.184147 -2.472115
H 2.773442 -1.642779 -2.162886
H 2.167541 -0.425827 -3.295768
O 1.836475 0.322827 1.507375
H 2.466396 0.044929 0.832848
C 2.321431 -0.011289 2.813605
H 3.260421 0.495421 3.016550
H 1.574876 0.324149 3.523141
H 2.453913 -1.085172 2.912730
O 0.760655 2.269667 -0.187975

H 1.257887 2.190840 0.633986
C 0.323582 3.620543 -0.380787
H -0.331878 3.932839 0.427559
H 1.173184 4.294010 -0.446458
H -0.223777 3.645302 -1.315165

37
**Na+ 6 formamide**
Na 0.004651 -0.026830 -0.004906
C 1.491020 0.809304 2.533072
N 2.212621 0.926641 3.652906
O 1.589103 -0.130150 1.757586
H 0.792369 1.630114 2.356055
H 2.103824 1.716353 4.258073
H 2.882155 0.215237 3.884224
C -1.631689 -1.254033 2.275160
N -2.392805 -1.529770 3.340000
O -1.657478 -0.187126 1.679523
H -0.971302 -2.067749 1.966586
H -2.344091 -2.420675 3.793847
H -3.032604 -0.835570 3.681021
C -0.462066 -2.664468 -1.487679
N -0.514802 -3.921029 -1.942599
O -0.160674 -2.370096 -0.340555
H -0.715368 -1.903432 -2.229025
H -0.776103 -4.118125 -2.888735
H -0.302773 -4.679538 -1.320184
C 2.745303 -0.736533 -1.164831
N 3.828251 -0.998835 -1.904468
O 1.792923 -0.081302 -1.562029
H 2.772810 -1.154557 -0.155872
H 3.884250 -0.640718 -2.840689
H 4.589103 -1.535021 -1.535688
C -2.674347 0.880146 -1.171868
N -3.726302 1.250879 -1.910348
O -1.708435 0.279671 -1.618562
H -2.740640 1.157390 -0.116701
H -3.745350 1.026220 -2.888464
H -4.499502 1.737636 -1.501913
C 0.535757 2.814857 -1.014930
N 0.613453 4.128453 -1.253568
O 0.156789 2.337818 0.044549
H 0.845742 2.182802 -1.850655
H 0.943281 4.475581 -2.132593
H 0.355997 4.776921 -0.531736

**Na+ 4acetone**
Na -0.001701 -0.017877 -0.009847
O -1.190423 -1.795446 0.797006
C -1.561412 -2.400620 1.786683
C -2.447653 -3.599906 1.673691
C -1.152306 -1.977013 3.162368
H -0.357277 -1.241841 3.111732
H -2.021920 -1.546827 3.659860
H -0.847379 -2.840470 3.749178
H -1.869181 -4.483809 1.942797
H -3.268335 -3.531341 2.384432
H -2.822846 -3.701381 0.662651
O -0.761985 1.986492 -0.802273
C -0.997963 2.662692 -1.787351
C -0.716927 2.158922 -3.168192
C -1.582612 4.033560 -1.663376
H -1.909879 4.216777 -0.647204
H -0.820190 4.761134 -1.942088
H -2.407157 4.156993 -2.362115
H -1.669150 1.941334 -3.652647
H -0.229644 2.930045 -3.760493
H -0.111806 1.260409 -3.127781
O 1.078629 0.661554 1.886166
C 1.500608 1.606723 2.527982
C 2.231718 1.402776 3.816528
C 1.311213 3.016155 2.062239
H 0.600345 3.052615 1.244650
H 0.988705 3.647253 2.887205
H 2.276675 3.398318 1.729518
H 1.599124 1.752996 4.632329
H 2.466303 0.354901 3.958605
H 3.136359 2.006548 3.837063
O 0.866508 -0.924170 -1.919995
C 1.051768 -1.939720 -2.566255
C 1.790895 -1.906230 -3.865901
C 0.551981 -3.269304 -2.095303
H 0.081777 -3.808765 -2.914309
H 1.409184 -3.862467 -1.775738
H -0.136623 -3.143569 -1.267612
H 2.532493 -2.701401 -3.899062
H 1.082692 -2.100376 -4.671564
H 2.257364 -0.939784 -4.013099

37
**Mg2+ 6 methanol**
Mg 0.064675 0.005828 0.034969
O -0.718945 -1.666653 -0.906392

H -0.334703 -2.521279 -0.648925
C -1.844180 -1.899026 -1.799018
H -2.280310 -0.931480 -2.006776
H -1.493457 -2.356582 -2.716806
H -2.571275 -2.538085 -1.311210
O 1.617880 -0.018134 -1.340930
H 1.566002 -0.635776 -2.073541
C 2.752169 0.875640 -1.529920
H 3.343026 0.536229 -2.371337
H 2.385942 1.877935 -1.717030
H 3.354853 0.855255 -0.630721
O 1.178682 -1.307317 1.199189
H 2.083152 -1.498963 0.942914
C 0.872799 -1.960762 2.464703
H 1.735679 -2.522382 2.799693
H 0.620086 -1.203001 3.195795
H 0.036667 -2.632386 2.312255
O 0.871547 1.649661 1.004015
H 0.690920 2.521592 0.646271
C 1.823878 1.755054 2.101339
H 2.011574 2.799288 2.317132
H 1.395941 1.274775 2.972531
H 2.746954 1.265841 1.815498
O -1.034260 1.305491 -1.153221
H -1.859093 1.660915 -0.815130
C -0.850971 1.710635 -2.540864
H -1.698824 2.304312 -2.858861
H 0.055175 2.298561 -2.613900
H -0.776223 0.819942 -3.153347
O -1.480590 0.056839 1.412478
H -1.433410 0.683876 2.137203
C -2.677018 -0.763446 1.543407
H -3.249891 -0.432729 2.400524
H -3.269360 -0.651398 0.643423
H -2.378810 -1.796206 1.675855

37
**Mg2+ 6 formamide**
Mg -0.000198 0.000059 0.000018
C 1.900891 1.343484 -1.842000
N 2.379557 1.798340 -2.983900
O 0.832738 0.726689 -1.746368
H 2.514049 1.543947 -0.964331
H 3.253017 2.291266 -3.019444
H 1.872464 1.648222 -3.839577
C -1.901278 -1.343412 1.842035
N -2.379994 -1.798125 2.983970

O -0.833179 -0.726523 1.746395
H -1.872983 -1.647825 3.839663
H -3.253389 -2.291166 3.019518
H -2.514330 -1.544112 0.964348
C -2.251377 1.536466 -1.179921
N -3.460296 1.756458 -1.659648
O -1.876332 0.438671 -0.750290
H -3.726901 2.663807 -1.995493
H -4.135690 1.011541 -1.687541
H -1.583727 2.397103 -1.180644
C 0.398186 2.153498 2.003470
N 0.496089 3.382728 2.471850
O 0.149773 1.891177 0.820220
H 0.695114 3.553392 3.440582
H 0.363592 4.168842 1.858407
H 0.549964 1.360411 2.734489
C -0.398801 -2.153258 -2.003430
N -0.496736 -3.382439 -2.471936
O -0.150181 -1.891067 -0.820194
H -0.364116 -4.168617 -1.858602
H -0.695817 -3.553001 -3.440674
H -0.550610 -1.360093 -2.734357
C 2.251009 -1.536401 1.179831
N 3.459988 -1.756426 1.659389
O 1.875939 -0.438599 0.750242
H 3.726603 -2.663773 1.995232
H 4.135403 -1.011525 1.687198
H 1.583329 -2.397015 1.180668

41
**Mg2+ 4 acetone**
Mg -0.004788 -0.006778 -0.003995
O -1.293924 -1.224268 0.778570
C -2.000748 -1.800573 1.613724
C -3.059457 -2.739651 1.177054
C -1.812245 -1.570808 3.066274
H -1.161052 -0.726075 3.255819
H -2.775803 -1.440427 3.554186
H -1.372837 -2.475486 3.491296
H -3.067494 -3.624638 1.809185
H -4.020772 -2.243648 1.329868
H -2.951855 -2.999988 0.131583
O -1.028773 1.459643 -0.756355
C -1.515042 2.184928 -1.632267
C -1.138522 2.022956 -3.057367
C -2.493873 3.236385 -1.272386
H -2.505808 3.416814 -0.204855

H -2.295354 4.149410 -1.828703
H -3.479270 2.888896 -1.591688
H -2.028966 2.036393 -3.682669
H -0.546883 2.892945 -3.348330
H -0.562497 1.119234 -3.216726
O 1.261121 0.630397 1.317091
C 1.965179 1.395892 1.986262
C 2.969347 0.854357 2.930578
C 1.828617 2.866990 1.862909
H 1.212640 3.141210 1.015023
H 1.372662 3.237537 2.783129
H 2.811023 3.330476 1.800724
H 2.971397 1.426330 3.855377
H 2.809364 -0.200232 3.116531
H 3.953411 0.996763 2.477640
O 1.032918 -0.896294 -1.382861
C 1.541071 -1.836503 -2.005298
C 2.414375 -1.574988 -3.172179
C 1.292709 -3.242706 -1.605649
H 1.119888 -3.860957 -2.483535
H 2.205048 -3.619047 -1.138088
H 0.468210 -3.317320 -0.906406
H 3.280129 -2.233145 -3.152699
H 1.850464 -1.836594 -4.070644
H 2.710840 -0.534704 -3.219384

25
**Mg2+ 4 acetonitrile**
Mg -0.030287 -0.011089 -0.023002
N -1.369999 -1.179053 1.012821
C -2.114796 -1.827656 1.588622
C -3.057155 -2.648303 2.317001
H -2.869274 -2.544931 3.382589
H -4.067244 -2.319414 2.086753
H -2.929861 -3.685761 2.018753
N -1.053826 1.365898 -1.158026
C -1.622260 2.131279 -1.788936
C -2.341489 3.099659 -2.587063
H -2.033665 4.100265 -2.294647
H -2.112836 2.935271 -3.636896
H -3.408274 2.977571 -2.418151
N 1.194353 0.970710 1.306665
C 1.875159 1.516389 2.045392
C 2.736590 2.206846 2.979895
H 3.768270 1.922823 2.789021
H 2.618321 3.279130 2.846962
H 2.458335 1.927658 3.992859

N 1.109342 -1.202477 -1.253230
C 1.742579 -1.864723 -1.936902
C 2.543792 -2.702604 -2.801819
H 2.372844 -2.412885 -3.835411
H 3.593239 -2.571956 -2.550706
H 2.256177 -3.740774 -2.657116

37
**Ca2+ 6 methanol**
Ca 0.000037 -0.001074 -0.001348
O -1.640774 -0.063226 1.708606
H -2.052905 -0.895420 1.957518
C -2.163690 0.988537 2.566558
H -3.236884 1.068969 2.437812
H -1.921697 0.774511 3.601511
H -1.683619 1.908048 2.256832
O -1.789792 0.275609 -1.530976
H -2.696650 0.320714 -1.214848
C -1.792064 0.454646 -2.974486
H -2.373171 -0.331826 -3.441972
H -2.200890 1.427124 -3.224714
H -0.760111 0.393237 -3.295672
O -0.333748 -2.336437 -0.233315
H -1.017526 -2.673117 -0.819374
C 0.378825 -3.464759 0.345321
H 0.846777 -4.047361 -0.440240
H 1.135325 -3.050670 0.999648
H -0.307452 -4.080475 0.915138
O 1.640825 0.061099 -1.711321
H 2.053001 0.893290 -1.960165
C 2.163688 -0.990626 -2.569352
H 3.236865 -1.071187 -2.440541
H 1.683496 -1.910119 -2.259760
H 1.921789 -0.776461 -3.604297
O 1.789856 -0.277745 1.528296
H 2.696719 -0.322836 1.212177
C 1.792113 -0.456824 2.971801
H 2.373228 0.329626 3.439314
H 0.760158 -0.395407 3.292979
H 2.200920 -1.429316 3.222004
O 0.333867 2.334286 0.230597
H 1.017587 2.670951 0.816732
C -0.378598 3.462624 -0.348143
H -0.846620 4.045257 0.437353
H 0.307769 4.078304 -0.917890
H -1.135036 3.048550 -1.002550

**Ca2+ 6 formamide**
Ca -0.075175 0.110750 -0.083129
C 1.894148 0.090418 2.595015
N 3.035177 -0.015163 3.254521
O 1.791140 -0.027512 1.371410
H 1.020886 0.294246 3.215747
H 3.067614 0.088221 4.251953
H 3.889013 -0.206337 2.759612
C -2.093432 -0.830631 2.391967
N -2.839383 -0.609370 3.460660
O -1.318806 0.002435 1.913538
H -3.456599 -1.315323 3.817614
H -2.804721 0.284297 3.920865
H -2.199829 -1.819091 1.944354
C -0.173985 -3.120341 -0.864382
N -0.524373 -4.387016 -0.720183
O -0.522069 -2.217103 -0.099811
H -0.214357 -5.086588 -1.369084
H -1.105922 -4.665708 0.050971
H 0.461537 -2.913896 -1.726343
C 2.785546 -0.418263 -1.703021
N 3.513014 -0.604870 -2.790885
O 1.551785 -0.445265 -1.691681
H 3.061685 -0.764467 -3.675462
H 4.515350 -0.575512 -2.755858
H 3.350588 -0.237460 -0.788253
C -2.982079 0.684549 -1.986766
N -3.256465 0.659682 -3.279407
O -1.886488 0.389455 -1.508577
H -2.547494 0.387969 -3.938546
H -4.167899 0.906641 -3.619181
H -3.812576 0.991801 -1.350476
C 0.524726 3.553465 -0.517350
N 0.178926 4.641909 0.147887
O 0.249710 2.408599 -0.157178
H 0.429732 5.554040 -0.187194
H -0.342288 4.563340 1.004099
H 1.086436 3.738569 -1.433443

**Ca2+ 4 acetone**
Ca 0.003062 -0.004002 0.018012
O -1.485012 -1.205693 1.200815
C -2.301880 -1.865351 1.850118
C -3.172482 -2.867207 1.185564
C -2.442545 -1.676202 3.315335

```
H -1.915811 -0.792801 3.653958
H -3.494598 -1.634039 3.589257
H -2.032042 -2.560158 3.807076
H -3.232557 -3.773576 1.783533
H -4.182719 -2.454590 1.147984
H -2.833380 -3.082639 0.179891
O -1.138800 1.512064 -1.187543
C -1.756974 2.329430 -1.875753
C -1.623777 2.337873 -3.353952
C -2.655706 3.329160 -1.246733
H -2.500909 3.383270 -0.176329
H -2.525441 4.302890 -1.713158
H -3.684084 3.024159 -1.450916
H -2.601820 2.442633 -3.818736
H -1.053848 3.226076 -3.632927
H -1.115256 1.451117 -3.711288
O 1.470505 1.023709 1.374423
C 2.275676 1.593159 2.117275
C 3.187029 0.814747 2.992615
C 2.360422 3.074186 2.157019
H 1.807711 3.525235 1.342432
H 1.946269 3.406757 3.110715
H 3.401065 3.391272 2.147552
H 3.243135 1.267805 3.979667
H 2.886817 -0.223836 3.054831
H 4.190093 0.879142 2.566033
O 1.155192 -1.352844 -1.365765
C 1.780768 -2.112950 -2.110474
C 2.581488 -1.582440 -3.241997
C 1.752323 -3.581811 -1.899617
H 1.629574 -4.097325 -2.849537
H 2.727674 -3.880591 -1.510568
H 0.978956 -3.866698 -1.197153
H 3.538613 -2.095363 -3.302070
H 2.047165 -1.817958 -4.164660
H 2.716073 -0.510730 -3.164450

25
```
**Ca2+ 4 acetonitrile**
```
Ca -0.031133 -0.011271 -0.022148
N -1.590555 -1.370821 1.181609
C -2.336469 -2.020262 1.757692
C -3.280668 -2.841443 2.485849
H -3.115879 -2.714109 3.552363
H -4.291249 -2.534095 2.230713
H -3.132814 -3.882797 2.212579
N -1.221365 1.593172 -1.341551
```

C -1.790442 2.359781 -1.972842
C -2.509720 3.329311 -2.772156
H -2.188816 4.329518 -2.493799
H -2.296221 3.153814 -3.823084
H -3.575646 3.221044 -2.590470
N 1.397578 1.130314 1.522520
C 2.080782 1.676697 2.260511
C 2.944344 2.367889 3.194675
H 3.977079 2.095586 2.994422
H 2.816427 3.440083 3.072479
H 2.677599 2.078916 4.207700
N 1.293784 -1.396454 -1.455746
C 1.927009 -2.059325 -2.141058
C 2.728527 -2.898082 -3.007222
H 2.560068 -2.607410 -4.040657
H 3.777983 -2.770684 -2.755688
H 2.439780 -3.936059 -2.865570

37
**Al3+ 6 methanol**
Al 0.101557 -0.003512 0.040508
O -0.665988 -1.546830 -0.794538
H -0.288132 -2.407987 -0.574080
C -1.793997 -1.747187 -1.752992
H -2.436971 -0.884390 -1.660712
H -1.381892 -1.840497 -2.749134
H -2.316599 -2.648519 -1.459071
O 1.564343 -0.082625 -1.208726
H 1.555717 -0.783178 -1.872369
C 2.681224 0.859722 -1.516603
H 3.404249 0.321323 -2.115546
H 2.276259 1.703187 -2.060771
H 3.118544 1.153108 -0.573407
O 1.109556 -1.187648 1.164600
H 2.045647 -1.299610 0.952957
C 0.792048 -2.005202 2.372407
H 1.128613 -3.016738 2.184696
H 1.304978 -1.571139 3.220491
H -0.279650 -1.966360 2.496694
O 0.865947 1.543267 0.887197
H 0.840236 2.383489 0.411863
C 1.641470 1.726855 2.149700
H 1.394262 2.703810 2.544677
H 1.319881 0.947967 2.825956
H 2.696638 1.649793 1.921396
O -1.327894 0.115530 1.315222
H -1.280948 0.838211 1.954457

C -2.585087 -0.651947 1.563031
H -2.676455 -0.793557 2.632416
H -3.415405 -0.079740 1.170135
H -2.475305 -1.598574 1.056575
O -0.876091 1.176852 -1.118965
H -1.631313 1.642874 -0.737427
C -0.800672 1.516289 -2.570576
H -1.787129 1.380489 -2.994953
H -0.471840 2.543397 -2.660189
H -0.093522 0.831142 -3.013655

37
**Al3+ 6 formamide**
Al 0.000349 -0.001019 -0.000359
C 1.879216 1.317674 -1.733789
N 2.310004 1.716539 -2.894930
O 0.800398 0.685733 -1.585842
H 2.500240 1.556274 -0.875779
H 3.179153 2.219964 -2.978869
H 1.788201 1.528370 -3.737823
C 0.400181 2.044341 1.982788
N 0.455859 3.280474 2.385169
O 0.146655 1.724897 0.791687
H 0.660258 3.502662 3.346800
H 0.291703 4.044594 1.746930
H 0.584976 1.279796 2.731176
C 2.139441 -1.536145 1.157654
N 3.354031 -1.693435 1.596983
O 1.714332 -0.441948 0.702622
H 3.663679 -2.583701 1.953957
H 4.015275 -0.931310 1.586487
H 1.490279 -2.405449 1.198811
C -0.399475 -2.046452 -1.983442
N -0.455036 -3.282599 -2.385796
O -0.145918 -1.726957 -0.792360
H -0.290804 -4.046686 -1.747537
H -0.659483 -3.504832 -3.347406
H -0.584386 -1.281943 -2.731838
C -1.878529 -1.319702 1.733067
N -2.309349 -1.718557 2.894201
O -0.799700 -0.687783 1.585127
H -1.787531 -1.530441 3.737096
H -3.178538 -2.221914 2.978130
H -2.499595 -1.558193 0.875057
C -2.138782 1.534110 -1.158260
N -3.353353 1.691365 -1.597648
O -1.713613 0.439874 -0.703373

H -3.663071 2.581684 -1.954429
H -4.014523 0.929173 -1.587370
H -1.489725 2.403509 -1.199131

41
**Al3 + 4 acetone**
Al -0.001402 -0.017928 -0.010067
O -1.198896 -1.156943 0.599816
C -1.947157 -1.717564 1.457910
C -2.923337 -2.710465 1.005262
C -1.840673 -1.378489 2.881237
H -1.269031 -0.475675 3.059265
H -2.833026 -1.318289 3.327163
H -1.352915 -2.228077 3.373375
H -3.020722 -3.514148 1.733093
H -3.897060 -2.202950 0.996738
H -2.711101 -3.080668 0.009922
O -0.912513 1.358271 -0.625861
C -1.509631 2.070535 -1.489886
C -1.472092 1.712852 -2.912128
C -2.240979 3.258345 -1.045134
H -1.954228 3.575983 -0.050341
H -2.154353 4.060259 -1.776170
H -3.303377 2.980666 -1.038151
H -2.449308 1.876717 -3.365361
H -0.801773 2.429478 -3.401081
H -1.117046 0.704108 -3.084961
O 1.118174 0.475256 1.257643
C 1.884034 1.258827 1.897876
C 2.681232 0.723521 3.003065
C 1.974097 2.678563 1.539187
H 1.535743 2.899345 0.573455
H 1.446374 3.236826 2.321323
H 3.009341 3.013881 1.594822
H 2.762808 1.452726 3.807256
H 2.318440 -0.235034 3.352645
H 3.700987 0.599508 2.615087
O 0.985641 -0.747647 -1.273838
C 1.572405 -1.681590 -1.901572
C 2.393202 -1.345666 -3.066415
C 1.431023 -3.076941 -1.471326
H 1.346252 -3.731697 -2.337727
H 2.374954 -3.354862 -0.987463
H 0.610769 -3.222944 -0.778752
H 3.269777 -1.989663 -3.115626
H 1.794730 -1.600177 -3.951127
H 2.655088 -0.295525 -3.102411

**Al3+ 4 acetonitrile**
 Al -0.028727 -0.010799 -0.024343
 N -1.244466 -1.068055 0.913594
 C -1.993465 -1.719268 1.493201
 C -2.927980 -2.532043 2.216145
 H -2.716428 -2.450131 3.283433
 H -3.939107 -2.179158 2.007546
 H -2.819932 -3.568259 1.892022
 N -0.955785 1.237843 -1.052330
 C -1.528314 2.006118 -1.687082
 C -2.244639 2.963480 -2.478793
 H -1.921816 3.967225 -2.197886
 H -2.029283 2.783095 -3.533143
 H -3.313253 2.848628 -2.290748
 N 1.080540 0.876574 1.183223
 C 1.764222 1.425272 1.926649
 C 2.616036 2.109615 2.855563
 H 3.649740 1.811583 2.673174
 H 2.506170 3.185379 2.710172
 H 2.322323 1.837977 3.870668
 N 1.004487 -1.090802 -1.138607
 C 1.641661 -1.756480 -1.825689
 C 2.436904 -2.585587 -2.684041
 H 2.244317 -2.308084 -3.721543
 H 3.490744 -2.433056 -2.446144
 H 2.164085 -3.629195 -2.519613

Table S2a Characterisation M (ion)..O or N distances in the investigated M..nL cluster. The distances are given in Å

|    | methanol | formamid | acetone | acetinitrlile | water |
|----|----------|----------|---------|---------------|-------|
| Li | 1,921    | 2,037    | 1,926   | 2,037         | 1,928 |
| Na | 2,4      | 2,372    | 2,285   | 2,396         | 2,386 |
| Mg | 2,073    | 2,061    | 1,931   | 2,051         | 2,086 |
| Ca | 2,372    | 2,273    | 2,248   | 2,393         | 2,369 |
| Al | 1,912    | 1,907    | 1,761   | 1,861         | 1,921 |

Table S2b Interaction energy in the investigated cluster m052x/cc-pVTZ level of theory (Eint=E(complex)-E(cat)-nE(mon), energy are given in kcal/mol)

| cc-pVTZ | methanol | formamid | acetone | acetinitrlile |
|---|---|---|---|---|
| Li | -121,86 | -142,80 | -125,28 | -132,30 |
| Na | -121,71 | -138,97 | -98,64 | -102,90 |
| Mg | -361,53 | -428,95 | -329,78 | -342,16 |
| Ca | -281,75 | -341,12 | -247,92 | -262,31 |
| Al | -792,02 | -924,18 | -748,24 | -780,78 |

Table S2c Interaction energy in the investigated cluster m052x/aug-cc-pVTZ/cc-pVTZ level of theory (E: kcal/mol)

| aug-cc-pVTZ | methanol | formamid | acetone | acetinitrlile |
|---|---|---|---|---|
| Li | -116,22 | -138,60 | -124,05 | -129,69 |
| Na | -112,92 | -133,53 | -97,60 | -100,46 |
| Mg | -353,27 | -422,22 | -328,83 | -339,57 |
| Ca | -279,29 | -340,57 | -252,74 | -266,74 |
| Al | -784,73 | -918,03 | -747,99 | -780,57 |

Table S2d Interaction energy in the investigated cluster m052x/cc-pVTZ level of theory using BSSE correction (E: kcal/mol)

| bsse/cc-pVTZ | methanol | formamid | acetone | acetinitrlile |
|---|---|---|---|---|
| Li | -115,27 | -137,99 | -120,94 | -129,78 |
| Na | -111,50 | -132,13 | -94,64 | -100,58 |
| Mg | -351,26 | -421,47 | -325,73 | -339,67 |
| Ca | -271,72 | -333,68 | -243,81 | -259,83 |
| Al | -780,42 | -915,63 | -746,13 | -778,02 |

Table S3 Mixing term in Magnesco- Perico scheme and Mayer bond order in the investigated M-nL complexes (n=4,6)

|        | acetonitrile | acetone | methanol | formamide | water |
|--------|--------------|---------|----------|-----------|-------|
| $Li^+$   | 0.22 | 0.17 | 0.18 | 0.16 | 0.04 |
| $Na^+$   | 0.14 | 0.10 | 0.08 | 0.08 | 0.03 |
| $Mg^{2+}$ | 0.30 | 0.26 | 0.16 | 0.18 | 0.04 |
| $Ca^{2+}$ | 0.17 | 0.12 | 0.05 | 0.11 | 0.11 |
| $Al^{3+}$ | 0.59 | 0.56 | 0.35 | 0.33 | 0.34 |
|        | acetonitrile | acetone | methanol | formamide | water |
| $Li^+$   | 0.34 | 0.38 | 0.37 | 0.36 | 0.36 |
| $Na^+$   | 0.19 | 0.19 | 0.16 | 0.22 | 0.19 |
| $Mg^{2+}$ | 0.45 | 0.50 | 0.33 | 0.35 | 0.33 |
| $Ca^{2+}$ | 0.30 | 0.32 | 0.28 | 0.29 | 0.29 |
| $Al^{3+}$ | 0.89 | 1.01 | 0.64 | 0.64 | 0.65 |

Fig. S1 Dipole moment of the investigated molecule taking into account only the geometrical deformation due to the cation. Ligand interaction.

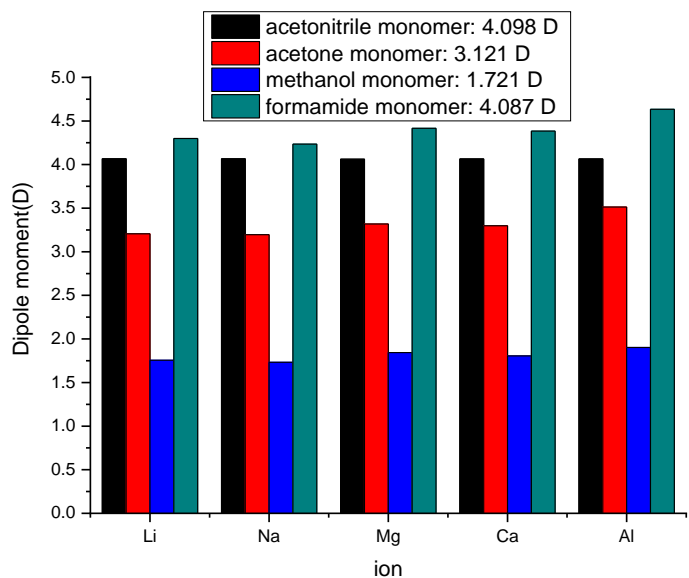